\documentclass[twocolumn]{aastex62}
\usepackage{natbib}
\usepackage{xspace}
\newcommand{\full}{``${\tt full1}$''\xspace}

\usepackage{xcolor}

\newcommand{\msunh}{\mathrm{M}_\odot h^{-1}}
\newcommand{\mpch}{{\rm Mpc}\,h^{-1}}

\newcommand{\kms}{{\rm km}\,\mathrm{s}^{-1}}

\defcitealias{2017MNRAS.467.1940H}{Hahn17}

\accepted{by ApJ}

\shorttitle{Fiber collision correction with MNN}
\shortauthors{Yang et al.}
\graphicspath{{figures/}}

\begin{document}

\title{Using the Modified Nearest Neighbor Method to Correct Fiber-collision Effects on Galaxy Clustering}

\correspondingauthor{Yipeng Jing}
\email{ypjing@sjtu.edu.cn}

\author{Lei Yang}
\affil{Department of Astronomy, School of Physics and Astronomy, Shanghai Jiao Tong University \\
Shanghai, 200240, China}

\affiliation{IFSA Collaborative Innovation Center, Shanghai Jiao Tong University \\
Shanghai, 200240, China}

\author{Yipeng Jing}
\affil{Department of Astronomy, School of Physics and Astronomy, Shanghai Jiao Tong University \\
Shanghai, 200240, China}

\affiliation{IFSA Collaborative Innovation Center, Shanghai Jiao Tong University \\
  Shanghai, 200240, China}

\affiliation{Tsung-Dao Lee Institute, Shanghai Jiao Tong University \\
Shanghai, 200240, China}

\author{Xiaohu Yang}
\affil{Department of Astronomy, School of Physics and Astronomy, Shanghai Jiao Tong University \\
Shanghai, 200240, China}

\affiliation{IFSA Collaborative Innovation Center, Shanghai Jiao Tong University \\
  Shanghai, 200240, China}

\affiliation{Tsung-Dao Lee Institute, Shanghai Jiao Tong University \\
Shanghai, 200240, China}

\author{JIAXIN Han}
\affil{Department of Astronomy, School of Physics and Astronomy, Shanghai Jiao Tong University \\
Shanghai, 200240, China}

\affiliation{Kavli IPMU (WPI), UTIAS, The University of Tokyo, Kashiwa, Chiba 277-8583, Japan}

\begin{abstract}
  Fiber collision is a persistent problem faced by modern spectroscopic galaxy surveys.
  In this work, we propose a new method to correct for this undesired
  effect, focusing on the clustering from the fiber-collision scale up to $\lesssim 10~\mpch$.
  We assume that the fiber-collided galaxies are in
  association with their nearest three angular
  neighbors.  Compared with the conventional nearest-neighbor method, we have properly accounted for
  the foreground (background) galaxies that are associated with the
  foreground (background) cosmic webs relative to the nearest
  neighbor. We have tested the new method with mock catalogs of the
  Sloan Digital Sky Survey (SDSS) Data Release 7 (DR7).  The test
  demonstrates that our new method can recover the projected two-point
  correlation functions at an accuracy better than 1\% on small 
  (below the fiber-collision scale) to intermediate
  (i.e., $10~\mpch$) scales, where the fiber collision takes effect and the
  SDSS main sample can probe. The new method also gives a better recovery of the
  redshift-space correlation functions almost on all scales that we
  are interested in. 
\end{abstract}


\section{Introduction} \label{sec:intro} 
The measurement of galaxy clustering plays a vital role in
observational cosmology. As one of the most powerful
probes of the growth rate of the matter density
field, it can put strong constraints on the many fundamental quantities, including 
the cosmological parameters~\citep{2002ApJ...564...15J, 2006ApJ...640L.119J, 
2006PhRvD..74l3507T, 2004PhRvD..69j3501T, 2004PhRvL..92x1302W, 
  2004MNRAS.350.1153Y, 2005PhRvD..71j3515S, 2012MNRAS.427.3435A, 
  2012MNRAS.425..415S, 2017MNRAS.464.1640S, 
  2013MNRAS.430..767C, 2013MNRAS.430..725V, 2014MNRAS.439.3504S,
   2018ApJ...861..137S}, neutrino masses, the
nature of gravity, and the properties of dark energy
\citep{1987MNRAS.227....1K, 2001Natur.410..169P, 2005ApJ...633..560E,
  2008Natur.451..541G, 2012MNRAS.426.2719R, 2013MNRAS.430..924C, 
  2017MNRAS.466.2242B, 2017arXiv170905173W}. At the same time,
galaxy clustering studies also provide crucial insights into the physics of
galaxy formation and their connections to dark matter halos on small and intermediate scales
\citep[e.g.,][]{2005Natur.435..629S, 2007MNRAS.374.1303C,2008ApJ...687...12D, 
2010gfe..book.....M, 2010ApJ...709..115W, 2013MNRAS.433..515W, 2018ApJ...852...31W, 
2014MNRAS.443.3044Z, 2015ApJ...806..125P, 2017ApJ...848...60Y}.

To accurately measure galaxy clustering, fiber-fed
spectroscopic galaxy surveys are generally required. However, these surveys often
come with the inevitable problem  of fiber collision~\citep{2006AJ....131.2332G, 2008ApJS..176..414Y,
  2013AJ....145...10D, 2015ApJS..219...12A, 2016MNRAS.455.1553R},
which arises from the fact that two fibers cannot be placed closer
than a separation limit called the fiber-collision scale. As a
result, a small fraction of galaxies in dense regions cannot be targeted for observation in these surveys.

For SDSS, the fiber-collision scale is
$55''$, resulting in $\sim 6\%$ of galaxies having no measured spectroscopic redshifts. 
The scale becomes slightly larger in 
the Baryon Oscillation Spectroscopic Survey (BOSS) \citep{2012MNRAS.427.3435A} and 
the Extended Baryon Oscillation Spectroscopic Survey 
\citep{2016AJ....151...44D}, which is $62''$ and the population without redshift is $\sim 5\%$. 
The case for the Dark Energy Spectroscopic Instrument 
(DESI) \citep{2013arXiv1308.0847L, 2016arXiv161100036D, 2016arXiv161100037D} is 
more complicated compared with other surveys, which is dedicated to 
completing the largest spectroscopic survey with a 5000-fiber spectroscopic instrument. 
By simulating the fiber assignment algorithm, \cite{2017JCAP...04..008P} 
found that only 49.5\% of luminous red galaxies (LRGs) and 11.6\% of 
the emission line galaxies (ELGs) can be observed for a one-pass survey 
in 1 deg$^2$, although the final expected achievements after multiple-pass observations 
for the full skies of LRGs and ELGs are 95\% and 78\%, respectively. 
Despite the fraction of fiber-collided galaxies being typically insignificant, its impacts on clustering are not trivial at all.
Recent studies gradually demonstrate that it not only affects the precise
clustering measurements below the fiber-collision scale, but also
biases measurements on intermediate and larger scales
\citep[e.g.,][]{2002ApJ...571..172Z, 2005ApJ...630....1Z, 2012ApJ...756..127G}.
Furthermore, the multipoles of the power spectrum measured in redshift space can
also be severely influenced by these effects (\citealt[][hereafter Hahn17]{2017MNRAS.467.1940H}). 
Correcting for this fiber-collision problem is thus crucial 
for the application of galaxy clustering in the era of precision cosmology.

So far, various methods have been proposed to correct for 
fiber-collision effects. They can be basically divided into two
categories. One category is to assign a redshift to 
each fiber-collided galaxy. For example, the nearest 
angular neighbor method simply assigns the redshift of the nearest neighbor 
to the fiber-collided galaxy~\citep{2005ApJ...630....1Z, 2011ApJ...736...59Z, 2006ApJS..167....1B}. 
Improved versions of this method are achieved by 
adding a distribution of the line-of-sight displacements between 
the fiber-collided galaxy and the nearest neighbor 
\citepalias{2017MNRAS.467.1940H}.  
The other category works by applying a weighting scheme to the pair counts 
in order to recover the true pair counts. The weights can be obtained from the 
angular correlation function~\citep{2003MNRAS.346...78H, 2006MNRAS.368...21L, 
2011ApJ...728..126W}, the redshift completeness of the observed 
galaxies~\citep{2012ApJ...756..127G}, the occurrence of close 
pairs~\citep{2012MNRAS.427.3435A, 2012MNRAS.424..564R, 
2017MNRAS.466.2242B, 2017MNRAS.465.1757G}, or the simulated selection function of the 
pairs~\citep{2017MNRAS.472.1106B, 2018arXiv180500951B}. 
Both categories have their own advantages and disadvantages. 
For the first category, the redshift assignment methods, though widely used, 
are unable to recover the true clustering below the fiber-collision scale. 
For instance, the line-of-sight reconstruction method of 
\citetalias{2017MNRAS.467.1940H} successful recovers the true power 
spectrum monopole on small scales compared with previous methods, 
but for the quadrupole power spectrum it shows little improvement. 
For the second category, the weighting algorithms 
generally require detailed tiling or spectroscopic mask information. 
These algorithms work very well in redshift space for high-completeness samples, 
but the correction to very low completeness samples may still show significant bias with 
large error bars. For example, by applying the method of \citet{2017MNRAS.472.1106B} to 
the VIMOS Public Extragalactic Redshift Survey \citep{2012PASP..124.1232G} mock catalogs 
with only $\sim$47\% completeness, \citet{2018A&A...610A..59M} found 
systematic underestimations of the multipole moments of the two-point 
correlation functions. Additionaly, the weighting methods are also difficult to implement in Fourier space.

Instead of relying on the observational data to correct for the fiber-collision effects,  
\citetalias{2017MNRAS.467.1940H} proposed an alternative approach to recover 
the true power spectrum. By modeling the fiber-collision effects through a convolution 
of the true power spectrum with a scaled top-hat function, their effective window method can model 
the fiber-collided power spectrum down to the scale of $k \approx0.83 h$Mpc$^{-1}$, both in monopole 
and quadrupole. However, the effectiveness of this method for recovering the 
two-point correlation functions in real space is yet to be tested.

In this paper, we introduce a new redshift assignment method to correct for the 
fiber-collision effect below the fiber-collision scale and the 
intermediate scale $\lesssim 10~\mpch$. The method falls into the first category, 
which can be used to measure clustering both in physical space and in Fourier space. 
We make an assumption that each fiber-collided galaxy is in association with its three
nearest angular neighbors, and the coherence length for each galaxy
pair is no more than 20 $\mpch$ \citep{2011ApJ...734...88W}. We test
our method with the two-point correlation statistics using mock
catalogs. Compared with the previous methods, our method recovers galaxy clustering with
smaller biases as well as smaller statistical errors on all scales.

The paper is organized as follows. We introduce the mock catalog
construction in Section \ref{sec:mock}.  We describe  our new method in detail, including 
its statistical basis in Section\ref{sec:method}. We present tests of the method and compare 
with some other methods in Section \ref{sec:test}.  Finally, we summarize in Section
\ref{sec:summary}.

\section{Construction of Mock Catalogs}\label{sec:mock}

We construct a mock galaxy catalog from a cosmological $N$-body simulation 
in the CosmicGrowth simulation suite~\citep{2018arXiv180706802J} $\rm WMAP\_3072\_600$. 
This simulation is performed with a parallel adaptive $\rm P^3M$
code adopting a standard flat $\Lambda \rm CDM $ cosmology. The parameters
are set as $\Omega_m=1-\Omega_\Lambda=0.268$, $\Omega_b=0.045$,
$h=H_0/(100\,\kms\mathrm{Mpc}^{-1})=0.71$, $\sigma_8=0.83$, and
$n_s=0.968$, which are compatible with the observations of 
the Nine-Year $Wilkinson~Microwave~Anisotropy~Probe$ ($WMAP$ 9;
\citealt{2013ApJS..208...20B,2013ApJS..208...19H}).  The simulation starts 
at an initial redshift of 144, and evolves with $3072^3$
particles in a cubic box of 600 $\mpch$ on a side, attaining a mass
resolution of $5.54\times 10^8 \msunh$.  There are 100
snapshots output evenly in the logarithm of the scale factor between
$z=16.9$ and $z=0$.  The friends-of-friends algorithm
\citep{1985ApJ...292..371D} is applied to find halos in each snapshot
with a linking length of 0.2 in units of the mean particle
separation. Then, subhalos are identified together with their merger 
history using the Hierarchical Bound-Tracing code
\citep{2012MNRAS.427.2437H,2018MNRAS.474..604H}.  We include halos containing at
least 50 particles in our halo catalog and pick the
snapshot at $z=0$ for mock catalog construction.

We take into account the existence of ``orphan'' galaxies in our mock catalog. 
As a subhalo orbits within its host halo, its mass gradually decreases due to 
tidal stripping from its host. In some cases the mass of the subhalo can be stripped 
to below the resolution limit of our subhalo catalog, while the galaxy residing 
at the center of the subhalo can still survive unless the subhalo has merged with its host halo. 
Such a galaxy is called an ``orphan'' galaxy~\citep{2004MNRAS.352L...1G,2011MNRAS.413..101G}. 
To identify ``orphan'' galaxies, we keep tracking the most bound particle of each subhalo 
whose mass has been stripped to below our minimum mass cut. For each of these ``orphan'' subhalos, 
we then compute an infall time, $t_{\rm infall}$, defined as the elapsed time from the epoch 
when the subhalo attains its peak mass during its evolution to the epoch of the analysis. 
We also estimate the expected time for the subhalo to merge into the center of its host halo, 
$t_{\rm merge}$, according to the fitting function of \citet{2008ApJ...675.1095J} (their Equation~(5)) 
with the factor $C=0.43$ and the orbital circularity $\epsilon=0.51$.  We keep all ``orphan''
subhalos satisfying $t_{\rm merge} > t_{\rm infall}$ in our halo catalog as
hosts of ``orphan'' galaxies. 

\begin{figure}
\plotone{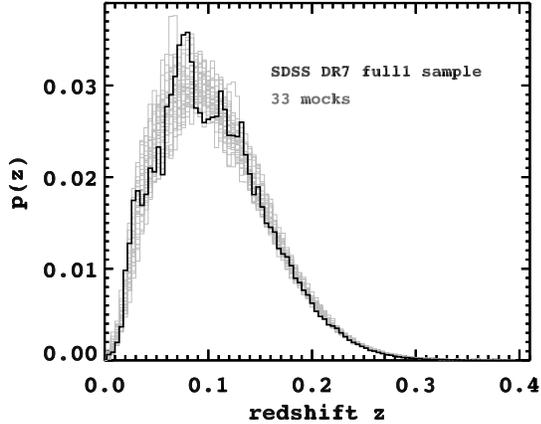}
\caption{Normalized galaxy redshift distribution of the SDSS DR7 \full
  sample (black) and that of the 33 mock galaxy samples (gray).  $p(z)$ is the fraction of
  galaxies in each redshift bin, with a bin width of $\Delta z=0.004$.\label{fig:zhist}}
\end{figure}

We use the subhalo abundance matching technique (
\citealt{2004ApJ...609...35K, 2006ApJ...647..201C,
  2006MNRAS.371.1173V, 2010ApJ...717..379B, 2010MNRAS.404.1111G, 2016MNRAS.459.3040G, 
  2012MNRAS.423.3458S, 2014MNRAS.437.3228G, 2016MNRAS.460.3100C,
  2018arXiv180403097W}) to link galaxies to their
host dark matter subhalos. Specifically, each subhalo in our catalog is matched to 
a galaxy with a given luminosity assuming a monotonic relation between the
galaxy absolute magnitude (or luminosity) and the peak mass of the subhalo,
$M_{\rm peak}$. Here, $M_{\rm peak}$ is the maximum mass that a subhalo 
ever had throughout its evolution history. We adopt the luminosity function of 
the SDSS DR7 \full sample as compiled from the New York University Value
Added catalog (NYU-VAGC)\footnote{http://sdss.physics.nyu.edu/lss/dr72/}
\citep{2001AJ....121.2358B,2003ApJ...592..819B,2005AJ....129.2562B}
to perform the matching.  The equation
\begin{equation}
n_g(>\mathrm{M}^{0.1}_{r})=n_{\rm subhalo}(>M_{\rm peak})
\end{equation}
is used to assign a galaxy magnitude, $\mathrm{M}^{0.1}_r$, to a subhalo, 
where $n_g$ and $n_{\rm subhalo}$ are the number densities of 
galaxies and subhalos, respectively. The galaxy is assumed to be located 
at the center of its assigned subhalo and inherits the position and velocity 
coordinates of the subhalo. We have not considered any scatter 
in the magnitude-mass correspondence 
$n$ in our matching. Adding a scatter to the relation should affect the
clustering of galaxies at the very luminous end. Since the aim of this
work is to test our method of correcting for the fiber-collision effect,
we believe adding the scatter should have little impact on our final
test results. After this step, we duplicate the simulation box periodically to create a
large box of mock galaxies. A random point within the box is then selected as the origin, 
and the galaxies are projected onto the celestial sphere to get their angular coordinates 
and true redshifts. We then derive the observed redshift of each galaxy, taking into account 
the peculiar velocity contribution, and the apparent magnitude, $m_r$, after $k-$ and $e-$corrections. 
After that, the galaxy catalog is trimmed by the MANGLE software \citep{2008MNRAS.385.1635S} 
according to the survey mask of the SDSS \full sample, with angular
and radial selection functions derived from NYU-VAGC. Finally, we use the
fiber-collision code \footnote{http://sdss4.shao.ac.cn/guoh/} of
\cite{2012ApJ...756..127G} to add fiber-collision effects to the
masked galaxy catalog. The fraction of fiber-collided galaxies in our
final mock catalog perfectly matches the fraction $\sim 5.6\%$ of the
SDSS \full sample. By shifting the origin and rotating the box, we
create a total of 33 mock catalogs for our following
two-point statistical analysis. Figure \ref{fig:zhist} shows the
normalized redshift distributions of the 33 mocks and that of the SDSS \full
sample. The agreement between the observed and the mock distributions is
remarkably good.

\section{The New Method} \label{sec:method}

In this section, we first analyze the statistical properties of galaxy pairs
using the galaxy population without fiber collision. Then, based on these
statistical properties, we elaborate our new approach to correct for the
fiber-collision effects.

Our new method is developed on top of the nearest-neighbor method and the method 
of \citetalias{2017MNRAS.467.1940H}. These previous methods make use of the distribution 
of the angular nearest neighbor to assign redshifts to fiber-collided galaxies. 
However, as we will see later, the nearest neighbor alone may not be sufficient to fully 
sample the redshift distribution of the fiber-collided galaxy. To improve over this, we also make use 
of the distribution of the second and third nearest neighbors in our redshift assignment scheme. 
Below we will present the statistical properties of these angular neighbors.

\subsection{Statistics of the Observed Galaxies} \label{sec:observation}

\begin{figure*} 
\gridline{\fig{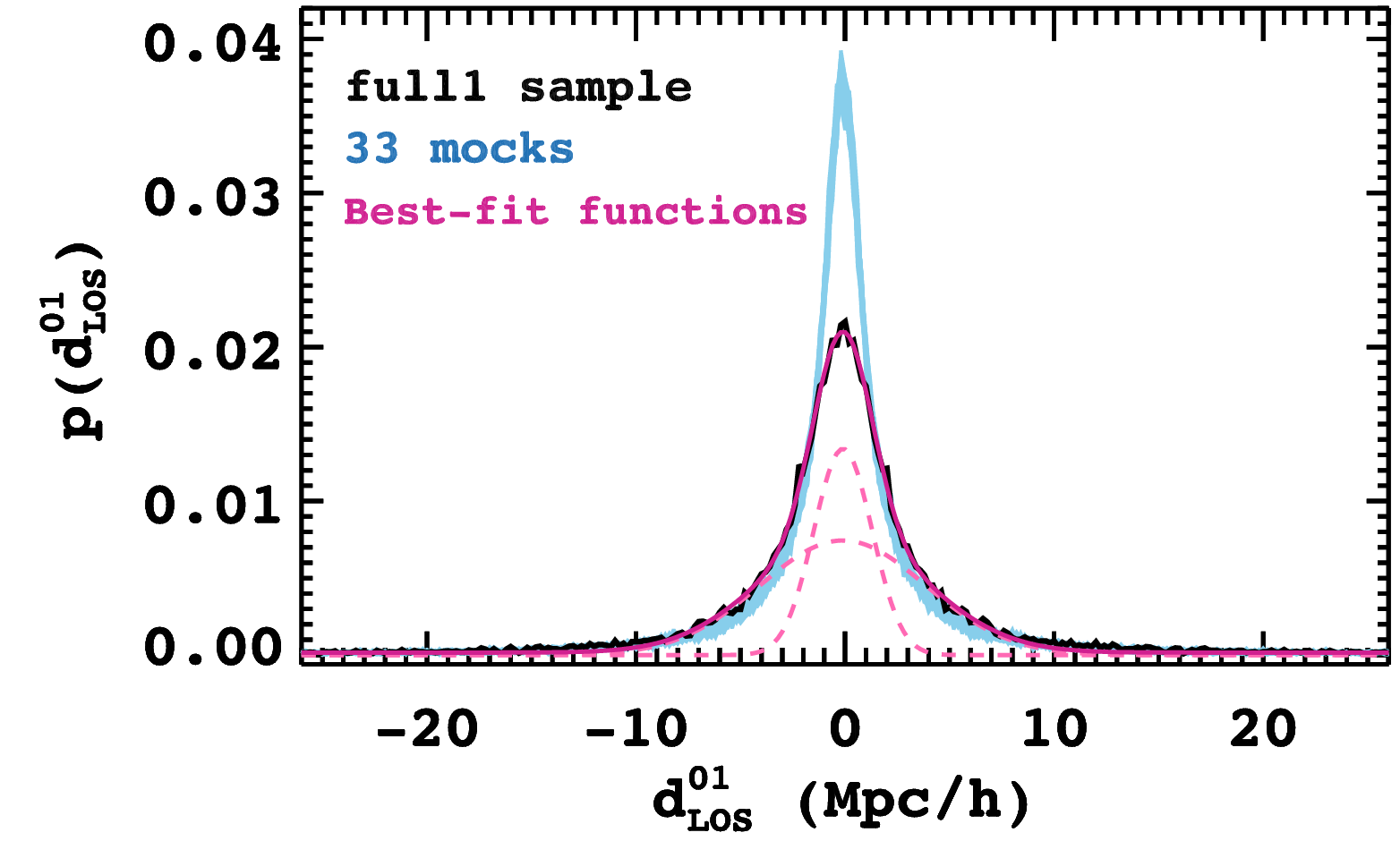}{0.56\textwidth}{(1)}}
\gridline{\fig{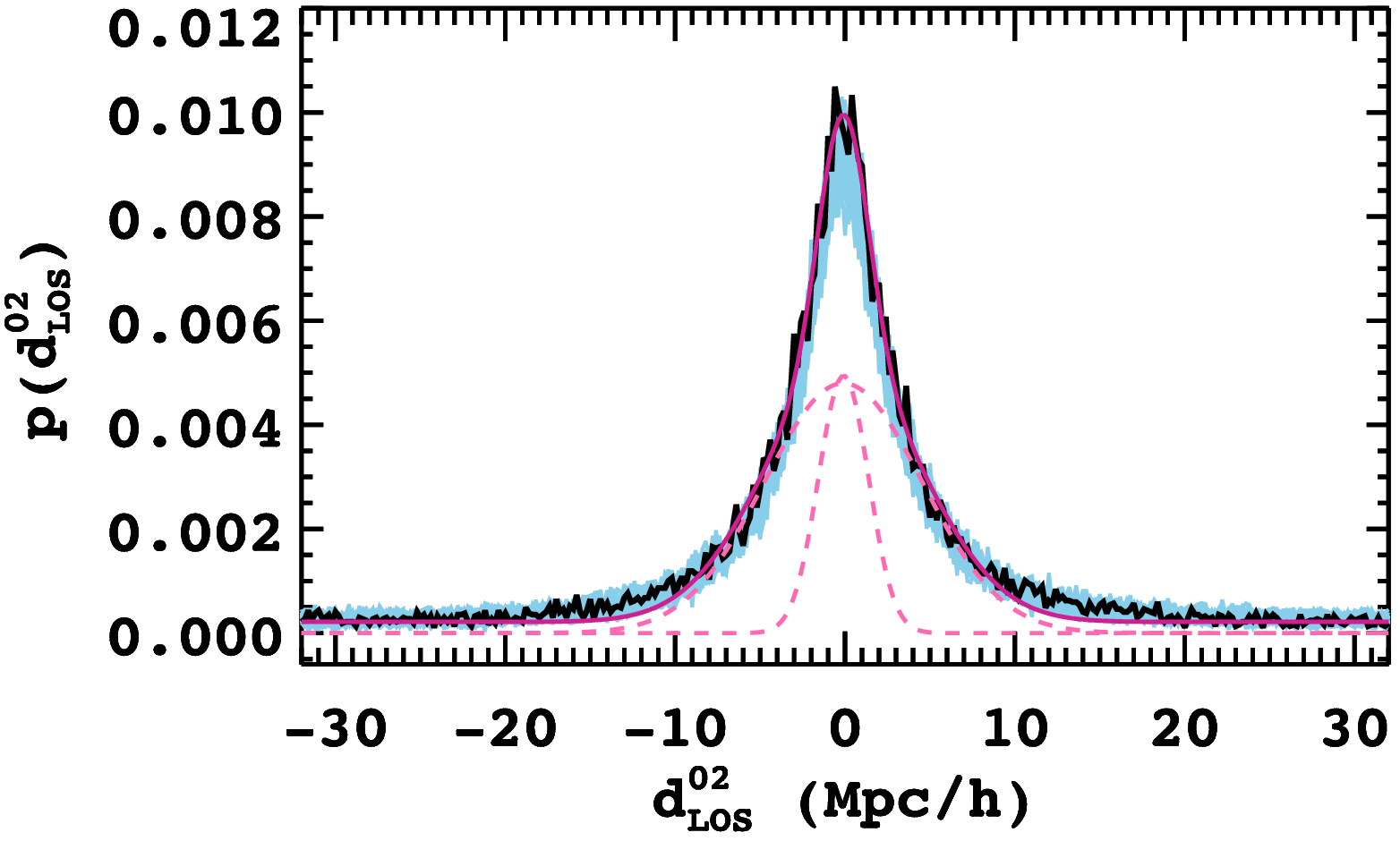}{0.48\textwidth}{(2)}
              \fig{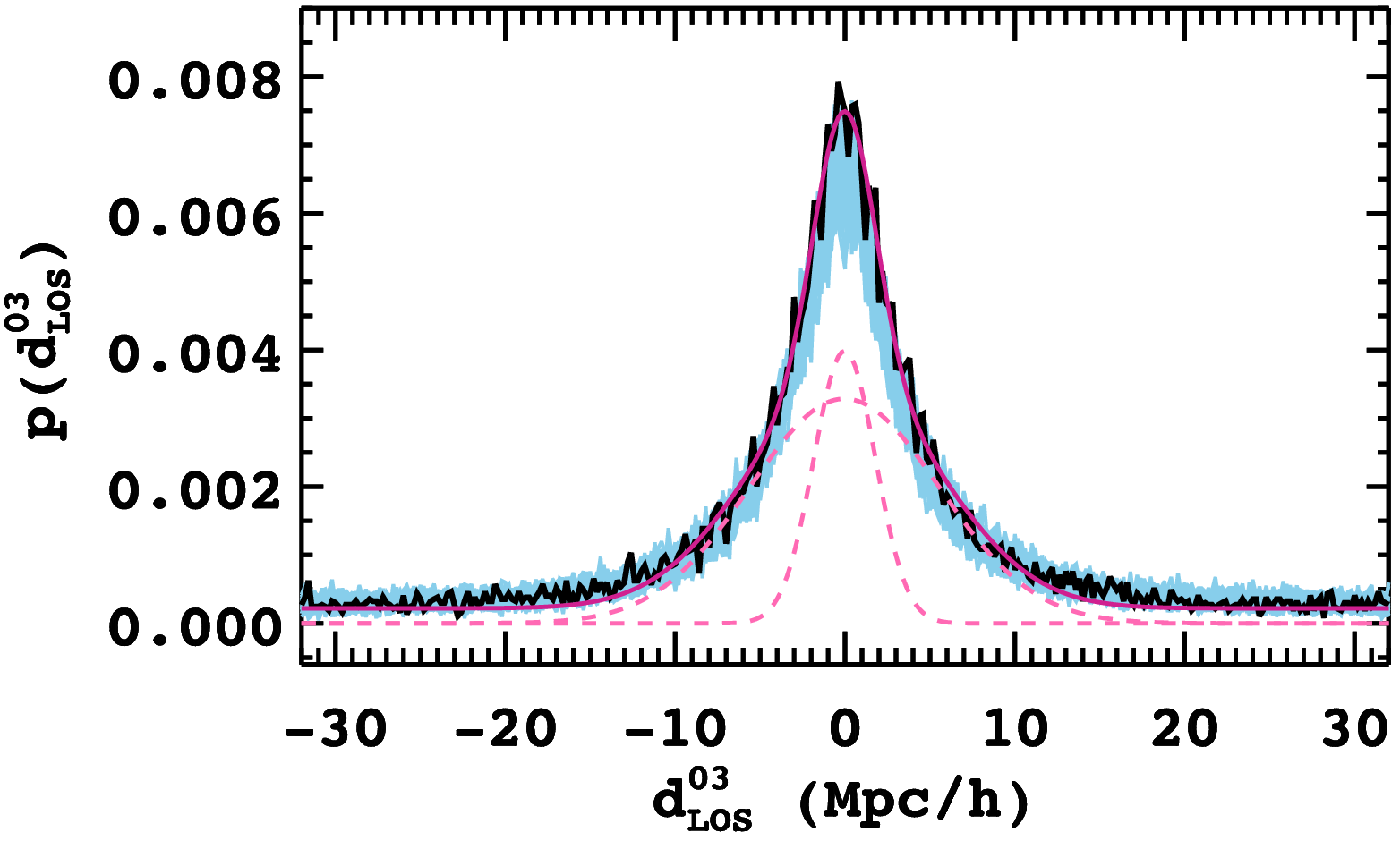}{0.48\textwidth}{(3)}
          }
          
          \caption{Normalized distribution of line-of-sight comoving separations between 
          a candidate fiber-collided galaxy ``0'' and its angular neighbors. 
          The three panels show the distribution of the separation to 
          the galaxy's nearest ($d_{\rm LOS}^{01}$), second nearest ($d_{\rm LOS}^{01}$), 
          and third nearest neighbors ($d_{\rm LOS}^{01}$) respectively. The bin sizes, 
          $\Delta d=0.2~\mpch$, are the same for different distributions.  
          The black histograms are for the SDSS DR7 \full\ sample, 
          while the blue histograms are distributions for the 33 mocks. 
          The violet red curves show fits to the \full sample histogram, 
          and the pink dashed curves show the two Gaussian components of each fit. \label{fig:hist}}
\end{figure*}

\begin{table}[t]
\begin{center}

\caption{The Best-fitting Parameters for Pair Distributions}\label{tab:bestfit}
{\small
\begin{tabular}{lcccccccccc}
\hline \hline

  & $d^{01}_{\rm LOS}$ \footnotesize{($\mpch$ )}  &$d^{02}_{\rm LOS}$   & $d^{03}_{\rm LOS}$  \\

\hline\hline
$\mu_1$ \footnotesize{($\mpch$ )}  &  -0.11     &    -0.10   &   -0.04  \\
$\sigma_1$  \footnotesize{($\mpch$ )}  &   3.98    &    4.64   &  5.58   \\
$A_1$   \footnotesize{($\mpch$)}  &  0.07       &  0.06     &   0.05   \\
$\mu_2$ \footnotesize{($\mpch$ )} &    -0.10     &   -0.09    &  -0.02    \\
$\sigma_2$ \footnotesize{($\mpch$ )}  &    1.39     &   1.47    &   1.81    \\
$A_2$  \footnotesize{($\mpch$)} &   0.05      &    0.02   & 0.02    \\
$B$   &    0.0002     &   0.0002    &  0.0002    \\
\hline\hline

\end{tabular}  
}
\end{center}
\tablecomments{\\
The best-fitting functions return three parameters for individual Gaussian functions in equation~(\ref{ffunc}) and a constant 
$B$. The three parameters are the mean $\mu$, $\sigma$, and the area $A$ of the best-fitting Gaussian curves. 
}

\end{table}

\begin{table*}[t]
\begin{center}
\caption{Statistics of galaxy pairs}\label{tab:dstatistics}
{\small
\begin{tabular}{lcccccccccc}
\hline \hline
 $d_{\rm LOS}$ \footnotesize{($\mpch$ )}    &   \full sample    &      33 Mocks  \\
\hline
     &  Fraction\tablenotemark{(a)}  &  Mean of fraction  ($1\sigma$) \\
\hline
$|d^{01}_{\rm LOS}| \le $ 20  &   66.2\%  & 71.5\%(0.40\%)\\
$|d^{02}_{\rm LOS}| \le $ 20  &    43.2\%  & 40.4\%(0.40\%)  \\
$|d^{03}_{\rm LOS}| \le $ 20  &   37.5\%  & 34.1\%(0.35\%) \\
$|d^{04}_{\rm LOS}| \le $ 20  &   34.2\% & 30.9\%(0.46\%)  \\
$|d^{05}_{\rm LOS}| \le $ 20  &   31.8\%  & 28.8\%(0.50\%) \\
\hline
$|d^{01}_{\rm LOS}|\le20$\&\tablenotemark{(b)} $|d^{02}_{\rm LOS}|  \le $ 20  &  34.4\% & 33.8\%(0.41\%) \\
$|d^{01}_{\rm LOS}|\le20$\& $|d^{02}_{\rm LOS}|  \le $ 20\&$|d^{03}_{\rm LOS}|\le20$  &  20.7\% & 18.7\%(0.38\%) \\
$|d^{01}_{\rm LOS}|\le20$\& $|d^{02}_{\rm LOS}|  \le $ 20\&$|d^{03}_{\rm LOS}|\le20$\&$|d^{04}_{\rm LOS}|\le20$  &  13.4\% & 11.4\%(0.33\%) \\
$|d^{01}_{\rm LOS}|\le20$\& $|d^{02}_{\rm LOS}|  \le $ 20\&$|d^{03}_{\rm LOS}|\le20$\&$|d^{04}_{\rm LOS}|\le20$\&$|d^{05}_{\rm LOS}|\le20$  &  9.2\% & 7.4\%(0.30\%) \\
\hline
$|d^{01}_{\rm LOS}|>20$\& $|d^{02}_{\rm LOS}|  \le $ 20  &  8.9\% & 6.7\%(0.14\%) \\
$|d^{01}_{\rm LOS}| > 20$\&$|d^{02}_{\rm LOS}| > 20$\&$|d^{03}_{\rm LOS}| \le $ 20   &  4.2\% & 3.3\%(0.08\%)  \\
$|d^{01}_{\rm LOS}| > 20$\&$|d^{02}_{\rm LOS}| > 20$\&$|d^{03}_{\rm LOS}| >20$\&$|d^{04}_{\rm LOS}| \le 20$   &  2.6\% & 2.2\%(0.07\%)  \\
$|d^{01}_{\rm LOS}| > 20$\&$|d^{02}_{\rm LOS}| > 20$\&$|d^{03}_{\rm LOS}| >20$\&$|d^{04}_{\rm LOS}| > 20$\&$|d^{05}_{\rm LOS}| \le 20$   &  1.9\%  & 1.6\%(0.06\%)  \\

\hline
$|d^{01}_{\rm LOS}|\le20$\&$|d^{02}_{\rm LOS}|  > $ 20  &  31.8\%  & 37.7\%(0.44\%)  \\
$|d^{01}_{\rm LOS}| \le 20$\&$|d^{02}_{\rm LOS}| \le 20$\&$|d^{03}_{\rm LOS}| > $ 20   &  13.7\% & 15.1\%(0.22\%)  \\
$|d^{01}_{\rm LOS}| \le 20$\&$|d^{02}_{\rm LOS}| \le 20$\&$|d^{03}_{\rm LOS}| \le20$\&$|d^{04}_{\rm LOS}| > 20$   &  7.3\% & 7.3\%(0.15\%) \\
$|d^{01}_{\rm LOS}| \le 20$\&$|d^{02}_{\rm LOS}| \le 20$\&$|d^{03}_{\rm LOS}| \le20$\&$|d^{04}_{\rm LOS}| \le 20$\&$|d^{05}_{\rm LOS}| > 20$   &  4.1\% & 4.0\%(0.18\%) \\

\hline\hline

\end{tabular}  
}
\end{center}

\tablecomments{\\
  $d^{01}_{\rm LOS}$ denotes galaxy pairs with $\Delta\theta^{01} \le 55''$ in galaxy samples, where 
  $\Delta\theta^{01}$ is the angular separation between the fiber-collided galaxy ``0'' and its first nearest-neighbor galaxy ``1'' in the spherical coordinate;\\
  (a) Fraction of galaxy pairs in total pairs that satisfy the corresponding condition. For example, there are 66.2\% of $d^{01}_{\rm LOS}$ pairs in the \full sample 
  have the comoving line-of-sight separations that are less than or equal to 20 $\mpch$, i.e., $|d^{01}_{\rm LOS}|  \le 20~\mpch$;\\
  (b) ``\&'' means multiple conditions are met at the same time.}  
\end{table*}

There are roughly $94.4\%$ galaxies with well measured spectroscopic
redshifts both in the SDSS DR7 \full sample and in our mock galaxy
catalogs. A small fraction of galaxy pairs within the fiber-collision
scale still have observed redshifts for both galaxies thanks to the overlapping tiling
regions in the survey. \citetalias{2017MNRAS.467.1940H} measured the line-of-sight comoving
distance separations of these close pairs with observed redshifts in BOSS DR12 CMASS. 
They found $70\%$ of the galaxy pairs
have $|d_{\rm LOS}| <  20~\mpch$, and the distribution of $d_{\rm LOS}$ can be roughly fitted 
by a Gaussian function within this distance range. The rest of the fiber-collided galaxy pairs, 
showing a flat ``tail'' extending to $\sim 500~\mpch$, follow a roughly uniform distribution.  In our
case, the fiber size of SDSS DR7 is $55''$ and the median redshift of
the \full\ sample is around $\sim 0.1$. These lead to a fiber-collision
scale of $0.1~\mpch$, which is smaller than the fiber-collided scale
of BOSS. In this work, we collect all 
galaxy pairs with an angular separation satisfying 
$\Delta\theta^{01} \le 55''$ and treat one of them as a 
pseudo-fiber-collided galaxy ``0.'' We measure, 
$d^{01}_{\rm LOS}=|d^0_{\rm LOS}-d^1_{\rm LOS}|$, the
line-of-sight comoving separation distance of the pseudo-fiber-collided galaxy
``0'' and its first angular nearest-neighbor galaxy ``1'' for all such 
galaxy pairs. Additionally, we also measure $d^{02}_{\rm LOS}=|d^0_{\rm LOS}-d^2_{\rm LOS}|$ 
and $d^{03}_{\rm LOS}=|d^0_{\rm LOS}-d^3_{\rm LOS}|$, the line-of-sight separation of galaxy ``0'' and its
second angular nearest-neighbor galaxy ``2'' and third angular nearest-neighbor 
galaxy ``3.'' Note these second and third nearest neighbors are not necessarily inside 
the fiber-collision scale from the pseudo-fiber-collided galaxy ``0.'' Figure \ref{fig:hist} shows the normalized 
distributions of $d^{01}_{\rm LOS}$, $d^{02}_{\rm LOS}$, and $d^{03}_{\rm LOS}$, respectively. 
These distributions can all be well fit by sums of two Gaussian functions with different 
parameters~\citep{2009ASPC..411..251M} as shown in Figure~\ref{fig:hist}. 
The best-fitting values for our two Gaussian functions,
\begin{equation}\label{ffunc}
p(x)=\frac{A_1}{\sigma_1\sqrt{2\pi}}e^{-\frac{(x-\mu_1)^2}{2\sigma_1^2}}+\frac{A_2}{\sigma_2\sqrt{2\pi}}e^{-\frac{(x-\mu_2)^2}{2\sigma_2^2}}+B,
\end{equation}
are presented in Table~\ref{tab:bestfit}, where $x$ refers to $d^{01}_{\rm LOS}$,
$d^{02}_{\rm LOS}$, or $d^{03}_{\rm LOS}$.
\footnote{Here, we find that a single Gaussian function does not give a very good fit to the pair distributions}

Table~\ref{tab:dstatistics} presents the fractions of galaxy pairs with line-of-sight separations 
less than or equal to or larger than 20 $\mpch$ in multiple conditions for the \full sample and 33 mocks. 
The superscript ``0'' denotes the central galaxy, and ``1'', ``2'', ``3'', ``4,'' and ``5'' represent the first, second, third, 
fourth, and fifth nearest angular neighbors (from near to far) of the central galaxy, respectively. 
$d_{\rm LOS}$ is the line-of-sight comoving separation between these galaxy neighbor pairs. 
For $d^{01}_{\rm LOS}$, we also require that the angular separation must be less than the angular scale of 
the fiber collision, $\Delta\theta^{01} \le 55''$. From the table, we see that the fraction of pairs with 
$d_{\rm LOS} \le 20~\mpch$ rapidly decreases as it goes to the third neighbor, 
and only decreases slowly when it comes to the fourth and fifth neighbors. Although there are $\sim 34\%$ of 
$d^{04}_{\rm LOS}$ and $\sim 31.8\%$ of $d^{05}_{\rm LOS}$ galaxy pairs in the \full sample with 
$d^{04,05}_{\rm LOS} \le 20~\mpch$, the fraction of their nearer-neighbor pairs satisfying 
$d^{01,02,03}_{\rm LOS} \le 20~\mpch$ simultaneously is extremely low, 
implying a very small probability that the fourth and 
the fifth neighbors are associated with the central galaxy ``0'' and other neighbors. 
The fraction distributions demonstrate that the third nearest neighbor of the central galaxy can be 
basically treated as a critical point when one tries to make a simple estimate on how many 
nearest neighbors are associated with the central galaxy. Furthermore, 
the fraction distributions of galaxy pairs can be imprinted in the clustering strength of galaxy
correlation functions, particularly on small and intermediate
scales. See Section~\ref{sec:diss} for a further description. 
Even on a large scale, galaxy pairs cannot be treated as random
distributions besause individual galaxies are settled in structures like
filaments, sheets, or cosmic webs
\citep{1996Natur.380..603B, 2001ApJ...557..495P,2014MNRAS.441.2923C}. Therefore,
in order to properly recover the redshift distribution of
fiber-collided galaxies and the pair distributions of angular
neighborhood galaxies, these statistics should be carefully taken into
account.

\subsection{The Modified Nearest Neighbor Method} \label{sec:MNN}

Based on the nearest angular neighbor method (hereafter, NN method;
\citealt{2006ApJS..167....1B,2011ApJ...736...59Z}) that has been further developed by
\citetalias{2017MNRAS.467.1940H}, the new fiber-collision correction method 
we present in this paper 
is called the modified nearest angular neighbor method ( MNN method). 
One key assumption of the MNN method is that the fiber-collided
galaxies are tightly correlated with their three nearest angular neighbors.

First, using the statistical results derived from the galaxies with 
the well measured spectroscopic redshifts (see Section~\ref{sec:observation} for details), 
we construct three independent subsamples of galaxies:

\begin{itemize}
\item $\mathbf{ \Phi^{01} =\{g~|~\Delta\theta^{01} \le 55'' \} }$, where $\mathbf{g}$ denotes galaxies. 
$\Delta\theta^{01}$ is the angular separation between the central galaxy ``0'' and its first angular nearest-neighbor 
galaxy ``1.'' Galaxy pairs in this subsample are selected to be within the fiber-collision scale, $\Delta\theta^{01} \le 55''$, and with measured redshifts due to the overlapping tiling regions. Therefore, the line-of-sight comoving separation of this pair is available by $d^{01}_{\rm LOS} \equiv |d^0_{\rm LOS}-d^1_{\rm LOS}|$, where 
$d^0_{\rm LOS}$ and $d^1_{\rm LOS}$ denote the radial comoving distance of 
galaxy ``0'' and galaxy ``1,'' respectively.

\item $\mathbf{\Phi^{02} =\{g~|~d^{02}_{\rm LOS} \le 20}~\mpch \}$. 
The second subsample is composed of galaxies ``0'' whose 
line-of-sight comoving separation to its second angular 
nearest-neighbor ``2'' satisfies $d^{02}_{\rm LOS} \le20~\mpch$. 
Here, $d^{02}_{\rm LOS} \equiv |d^0_{\rm LOS}-d^2_{\rm LOS}|$, where 
$d^0_{\rm LOS}$ and $d^2_{\rm LOS}$ denote the radial comoving distance of 
galaxy ``0'' and its second angular neighbor galaxy ``2,'' respectively.

\item $\mathbf{\Phi^{03} =\{g~|~d^{03}_{\rm LOS} \le 20 }~\mpch \}$. 
The third subsample consists of galaxies ``0'' whose 
line-of-sight comoving separation to its third angular 
nearest-neighbor ``3'' satisfies $d^{03}_{\rm LOS} \le20~\mpch$. 
Here, $d^{03}_{\rm LOS} \equiv |d^0_{\rm LOS}-d^3_{\rm LOS}|$, where 
$d^0_{\rm LOS}$ and $d^3_{\rm LOS}$ denote the radial comoving distance of 
galaxy ``0'' and its third angular neighbor ``3,'' respectively.

\end{itemize}
Note that galaxy ``0'' in $\mathbf{\Phi^{01}}$, $\mathbf{\Phi^{02}}$, 
and $\mathbf{\Phi^{03}}$ may be different, depending on individual selection conditions. 
We also refer to galaxy ``0'' as 
the pseudo-fiber-collision galaxy, becasue some of these galaxies are below 
the fiber-collided scale but their redshifts are measured. 
 
For the fiber-collided population, the comoving distance
$\widetilde{d^0}_{\rm LOS}$ of the fiber-collided galaxy is missing, but the redshifts of its
angular neighbors have been well measured. For clarity, we use ``$\sim$'' to 
denote the fiber-collision galaxies and their neighbors, 
distinguishing from the pseudo-fiber-collision galaxies. 
For each fiber-collided galaxy ``$\widetilde{0}$,'' we find the three angular nearest 
neighbors to the galaxy ``$\widetilde{0}$'' in the fiber-collision-free population. 
We label the three neighbors ``$\widetilde{1}$,'' ``$\widetilde{2}$,'' ``$\widetilde{3}$'' 
according to the size of the angular separation satisfying 
$\Delta\widetilde{\theta^{01}} \le \Delta\widetilde{\theta^{02}} \le \Delta\widetilde{\theta^{03}}$. 
Because the neighbors are searched in the fiber-collision-free galaxies, 
the redshifts of these neighbors are also known. We use $\widetilde{d^1}_{\rm LOS}$, 
$\widetilde{d^2}_{\rm LOS}$, and $\widetilde{d^3}_{\rm LOS}$ to represent the 
comoving distance of the neighbor galaxies ``$\widetilde{1}$,'' ``$\widetilde{2}$,'' ``$\widetilde{3}$,'' 
respectively. The traditional nearest angular neighbor method sets 
$\widetilde{d^0}_{\rm LOS}=\widetilde{d^1}_{\rm LOS}$, overestimating the 
correlation of galaxy ``$\widetilde{0}$'' and its nearest neighbor ``$\widetilde{1}$.'' 
\citetalias{2017MNRAS.467.1940H} added a Gaussian distributed
displacement $ d^{01}_{\rm LOS}$ to the equation,
$\widetilde{d^0}_{\rm LOS}=\widetilde{d^1}_{\rm LOS} + d^{01}_{\rm LOS}$, 
corresponding to $\sim 70\%$ fiber-collided galaxies in their CMASS sample, 
then they kept the remaining $\sim 30\%$ galaxies as $\widetilde{d^0}_{\rm LOS}=\widetilde{d^1}_{\rm LOS}$. 
This method alleviates the strong bias of the monopole power spectrum compared with the NN method, 
but there is still an obvious bias of the quadrupole power spectrum shown in their Figure 6.

Our MNN method is primarily developed based on the NN method and the method of \citetalias{2017MNRAS.467.1940H}. 
Detailed steps are listed as follows:
\begin{enumerate}
\item We construct 11 bins of galaxy pairs according to the angular separations of the pairs, denoted by $\Delta\theta_i$, ranging from $0''$ to $55''$ with a step 
$\delta\Delta\theta = 5''$, where $i=$1 to 11. Both $\Delta\theta^{01}$ of the 
fiber-collision-free population and $\Delta\widetilde{\theta^{01}}$ of the 
fiber-collided population fall into one of these bins.

\item For each galaxy ``$\widetilde{0}$,'' if $\Delta\widetilde{\theta^{01}}$ 
belongs to the $j$th bin $\Delta\theta_j$, we find galaxies ``$0$'' in the subsample 
$\mathbf{ \Phi^{01} }$ with $\Delta\theta^{01}\in \Delta\theta_j$. 
Out of these galaxy ``$0$''s we select 30 galaxies whose line-of-sight distances $ d^0_{\rm LOS}$ are the closest to the distance $\widetilde{d^1}_{\rm LOS}$ of galaxy ``$\widetilde{1}$.''
After that, a galaxy is \textbf{\emph{randomly}} selected from the 30 galaxies, then 
the $d^{01}_{\rm LOS}$ of this galaxy is set as $\widetilde{d^{01}}_{\rm LOS}=d^{01}_{\rm LOS}$.
 
  Finally, the new comoving distance for the fiber-collided galaxy ``$\widetilde{0}$'' 
  is derived by $\widetilde{d^0}_{\rm LOS}=\widetilde{d^1}_{\rm LOS} +
  \widetilde{d^{01}}_{\rm LOS}$. If the new $\widetilde{d^0}_{\rm LOS}$ falls 
  out of the distance range of the whole sample
  [$d^{min}_{\rm LOS},d^{max}_{\rm LOS}$], we repeat the process of selecting one galaxy 
  from $N_{\rm near}$ galaxies until a good $\widetilde{d^{01}}_{\rm LOS}$ is
  obtained.

\item We keep $\widetilde{d^0}_{\rm LOS}$ for the fiber-collided galaxy if
  $\widetilde{d^{01}}_{\rm LOS}\le20~\mpch$. Since
  $\widetilde{d^{01}}_{\rm LOS}$ follows the distribution of
  $d^{01}_{\rm LOS}$, $\sim 71\%$ of the fiber-collided galaxies will get an assigned distance,
  $\widetilde{d^0}_{\rm LOS}$. For the remaining $29\%$ of galaxies, they are
  supposed to be located at a distance of at least 20 $\mpch$ away from
  their nearest neighbor. So, we \textbf{\emph{randomly}} select a galaxy 
  from $\mathbf{ \Phi^{02} }$ and set
  $\widetilde{d^{0'}}_{\rm LOS}=\widetilde{d^2}_{\rm LOS} +
  d^{02}_{\rm LOS}$. Note that we use the second nearest neighbor 
  of galaxy ``$\widetilde{0}$'' instead of the first one.
  
\item We keep $\widetilde{d^{0'}}_{\rm LOS}$ for the fiber-collided galaxy 
  in step 3 if $|\widetilde{d^{0'}}_{\rm LOS}-\widetilde{d^1}_{\rm LOS}| > 20~\mpch$. 
  For the case of $|\widetilde{d^{0'}}_{\rm LOS}-\widetilde{d^1}_{\rm LOS}| \le 20~\mpch$, 
  we \textbf{\emph{randomly}} select a galaxy from $\mathbf{ \Phi^{03} }$ and 
  set $\widetilde{d^{0}}_{\rm LOS}=\widetilde{d^3}_{\rm LOS} + d^{03}_{\rm
    LOS}$.  Note that we use the third nearest neighbor here.
\end{enumerate}
 
After the above steps, all fiber-collided galaxies are assigned new comoving distances.
We highlight that selecting $d^{01}_{\rm LOS}$ on the basis of
$\Delta\theta$ bins is equivalent to appropriately adding the angular weight
to the selection process, as neighbor pairs in the same angular bin should have  
similar angular correlation features. In step 2, different numbers of $N_{\rm near}$
would result in tiny changes of the clustering strength on
small scales. We test this with $N_{\rm near}=60$, $100$ in Section~\ref{sec:testptrs}, 
and both cases give negligible changes under the measurement errors. 
So, the choice of $N_{\rm near}$ has no significant 
effect on our clustering results. In steps 3 and 4, we assign galaxies 
with $\widetilde{d^{0}}_{\rm LOS} >  20~\mpch$ to the position
of their second neighbor (step 3) and third neighbor (step 4), naturally inheriting the intrinsic 
scatter of these pair distributions. In this way, these fiber-collided galaxies can 
be located in their second and third galaxy associations.

\section{Testing the New Method} \label{sec:test} 

In this section, we test the performance of the MNN method on 
recovering the projected two-point correlation functions (P2PCFs) and the multipole
moments of correlation functions in redshift space using mock galaxy
samples. In particular, we will compare the performance of our method with those of the NN method and the
\citetalias{2017MNRAS.467.1940H} method.

\subsection{Clustering Estimators} \label{sec:clustering}

Following the common way to calculate the correlation function
\citep{1988ASPC....4..257H,1992ApJ...385L...5H, 1994MNRAS.266...50F},
we define the redshift separation vector
$\mathbf{s} \equiv \mathbf{v}_1-\mathbf{v}_2 $ and the line-of-sight
vector $\mathbf{l}\equiv (\mathbf{v}_1+\mathbf{v}_2)/2$, where
$\mathbf{v_1}$ and $\mathbf{v_2}$ are the redshift-space position vectors of a pair of galaxies. 
The separations parallel ($\pi$) and perpendicular
($r_p$) to the line of sight are derived as
\begin{equation}
 \pi \equiv \frac{\mathbf{s}\cdot \mathbf{l}}{|\mathbf{l}|}, ~~~~~~r^2_p \equiv \mathbf{s}\cdot \mathbf{s}-\pi^2,
\end{equation}
and $\mu=\pi/s= \rm{cos}\theta$, where $\theta$ is the angle between
$\mathbf{s}$ and $\mathbf{l}$.  A grid of $\pi$ and $r_p$ is
constructed by taking 2 $\mpch$ as the bin size for $\pi$
linearly up to 60 $\mpch$ and 0.2 dex as the bin size for
$r_p$ logarithmically in the range of [0.01, 40] $\mpch$.  The
estimator of \cite{1993ApJ...412...64L} is adopted as
\begin{equation}
 \xi(r_p,\pi)  = \frac{DD-2DR+RR}{RR},
\end{equation}
where DD, DR, and RR are the numbers of data-data, data-random,
and random-random pairs weighted by the angular completeness. By
integrating $\xi(r_p,\pi)$ along the line-of-sight separation $\pi$ we
derive the P2PCF \citep{1983ApJ...267..465D},
\begin{equation}
w_p(r_p)\equiv 2\int^{\infty}_0 \xi(r_p,\pi)~d\pi = 2\int^{\pi_{max}}_0 \xi(r_p,\pi)~d\pi.
\end{equation}
We also calculate the non-zero multipole moments $\xi_0$, $\xi_2$,
$\xi_4$ of the redshift-space 2PCF $\xi(s,\mu)$, because these
quantities can be used to study the redshift distortion effects and
put crucial constraints on cosmological parameters and dark energy
models
\citep{1992ApJ...385L...5H, 2001Natur.410..169P, 2004PhRvD..70h3007S, 
2005ApJ...633..560E, 2012MNRAS.426.2719R, 2012MNRAS.425..415S, 
2013A&A...557A..54D, 2016ApJ...833..287L}. The
multipole expansion of $\xi(s,\mu)$ is \citep{1992ApJ...385L...5H}
\begin{equation}
\xi(s,\mu) = \sum_i\xi_l(s)P_l(\mu),
\end{equation}
where $P_l$ is the $l$th order Legendre polynomial. The multipole
moment $\xi_l$ can be calculated as
\begin{equation}
\xi_l(s) = \frac{2l+1}{2}\int^1_{-1}\xi(s,\mu)P_l(\mu)d\mu.
\end{equation}
We use the bin size for $s$ in the same way as for $r_p$, and $\mu$
is cut into 20 bins with a linear step $\Delta\mu=0.1$ in the range of
[-1,1]. Generally, we use random points of 30 to 50 times the number
of galaxies to reduce the shot noise. The $\tt KDTREE$ code of
\cite{2004physics...8067K} is applied to accelerate the calculation of
the pair counts.

\subsection{Correlation Functions} \label{sec:cf}

\begin{table*}[t]
\begin{center}
\caption{Galaxy samples }\label{tab:samples}
{\small
\begin{tabular}{lcccccccccc}
\hline \hline
$M^{0.1}_r$  & $z_{\rm min}$  & $z_{\rm max}$  &  $N_{\rm true}$  &  $N_{\rm MNN}$   &  $N_{\rm NN}$   & $N_{\rm Hahn}$\\
\hline\hline
$\le$ -19&  0.02 & 0.064   &  78898(5988)   & 78296(5889)    &  78373(5957)    &   78372(5957)\\
$\le$ -20&  0.02 &  0.106   & 134479(5645)  & 133853(5545)  &  134040(5624)  &   134039(5625) \\
$\le$ -21&  0.02 &  0.159    &  77382(1446)   &  77718(1444)   &  77821(1453)   &   77820(1454)\\
\hline\hline

\end{tabular}  
}
\end{center}

\tablecomments{Samples are constructed using galaxies with fgot$\ge 0.5$ and $m_r \le 17.77$. 
Galaxies are cut into three luminosity bins with different redshift ranges. 
$N$ is the mean number of luminosity bins, and the numbers in brackets are $1\sigma$ variation among the samples. 
For $true$,the  MNN method, the NN method, and the \citetalias{2017MNRAS.467.1940H} method, 33 mocks are used.}

\end{table*}

\begin{figure*}
\gridline{\fig{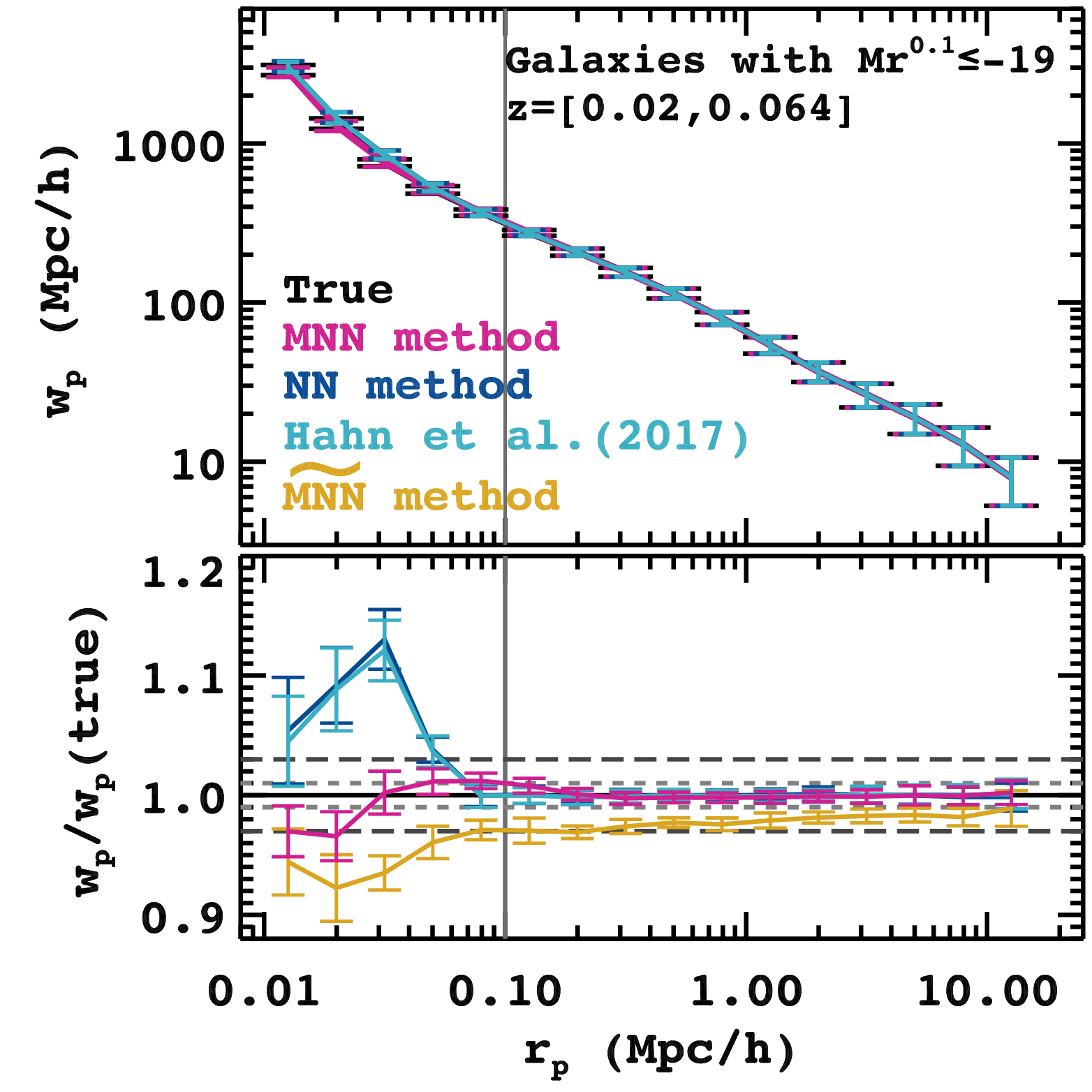}{0.33\textwidth}{(1)}
              \fig{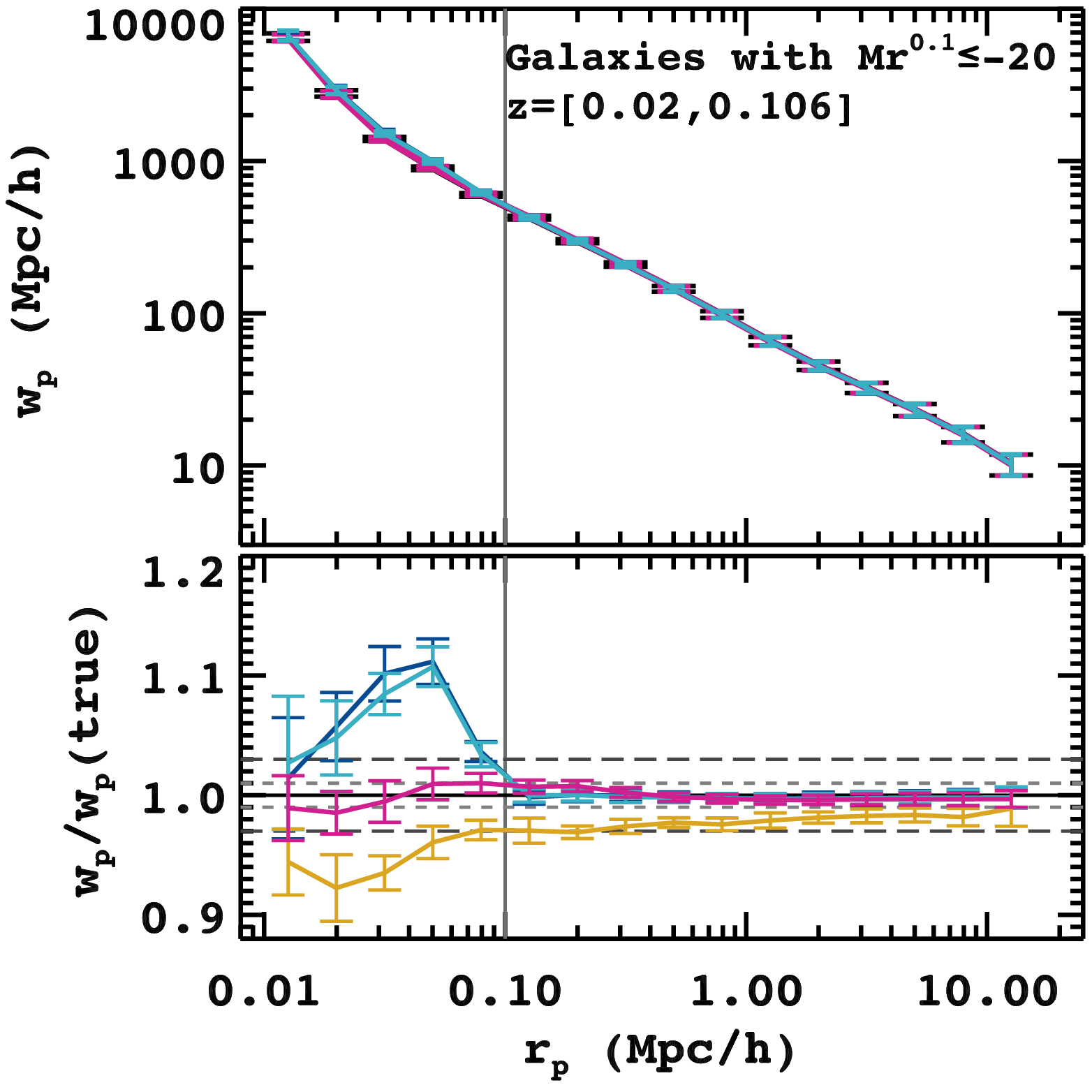}{0.33\textwidth}{(2)}
              \fig{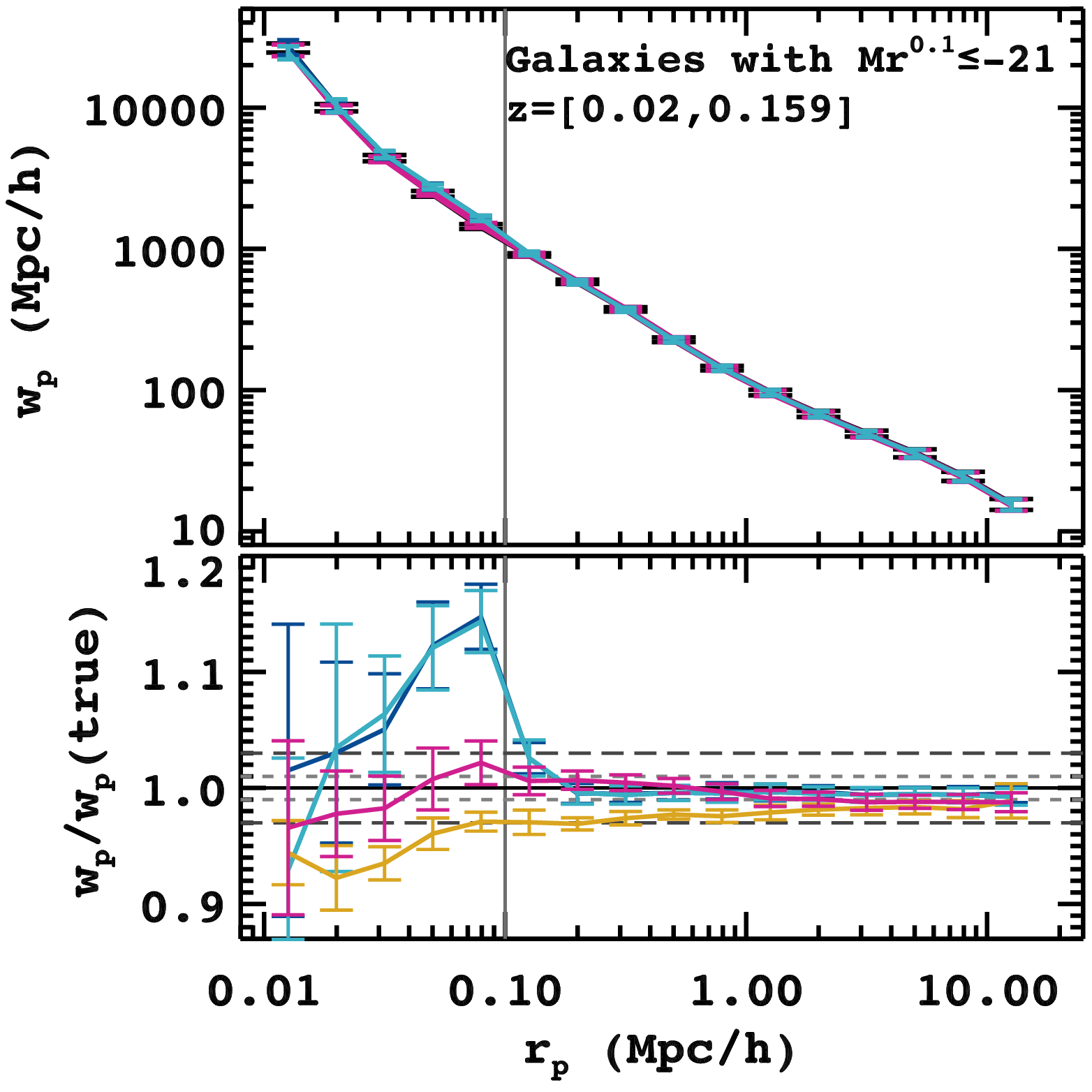}{0.33\textwidth}{(3)}
          }
          \caption{Comparisons of the projected two-point correlation
            functions $w_p(r_p)$ estimated from different fiber
            collision correction methods for volume-limited galaxy
            samples within three luminosity thresholds. In the upper
            panels, the redshifts of the fiber-collided galaxies are from
            the $true$ redshifts (black), redshifts generated from the MNN
            method (magenta), the NN method (dark blue), and the 
            \citetalias{2017MNRAS.467.1940H} method (light blue),
            separately.  Every $w_p(r_p)$ as shown in figures is the
            mean value of 33 mock samples for $w_p(true)$, 
            $w_p(\rm NN)$, and $w_p(\rm Hahn)$ method, and 99 mock samples for
            $w_p(\rm MNN)$.  Error bars of $w_p(r_p)$ are $1\sigma$
            variations of $w_p(r_p)$ among the samples. In the lower
            panels, the mean ratios of $w_p(r_p)$ estimated from
            different methods are presented, i.e.,
            $w_p(\rm MNN)$/$w_p(true)$ (magenta),
            $w_p(\rm NN)$/$w_p(true)$ (dark blue),
            $w_p(\rm Hahn)$/$w_p(true)$ (light blue), and
            $w_p(\rm \widetilde{MNN})$/$w_p(true)$ (yellow). To
            highlight the importance of introducing the second and
            third neighbors on fiber-collision corrections, we measure
            $w_p(\rm \widetilde{MNN})$ by just executing the first two
            steps of the MNN method. In this case $29\%$ of galaxy pairs
            are separated at $\widetilde{d^{01}}_{\rm LOS}>20\mpch$ following 
            step 2 with no further optimizations of the pair
            distributions.  The error bars for ratios are the $1\sigma$
            variation among the ratios of all samples belonging to the
            same volume-limited luminosity bin.  For clarity, we also mark the
            $1\%$ (horizontal short dashed gray lines) and $3\%$
            (horizontal dashed gray lines) bias levels in lower
            panels. The vertical straight gray lines
            denote the fiber-collision scale. \label{fig:wp}}
\end{figure*}

In order to explore the influence of the fiber-collision effect on clustering for
different luminous populations and to assess the performances of
those different methods, we construct three volume-limited galaxy samples
with different luminosity thresholds as shown in
Table~\ref{tab:samples} \citep{2015MNRAS.453.4368G}.  For the 
fiber-collided galaxies, after applying the MNN, NN, and the
\citetalias{2017MNRAS.467.1940H} methods, we reestimate M$^{0.1}_r$ with $k$-
and $e$-corrections based on the new redshifts and apparent magnitudes.
For the $true$ redshifts and redshifts derived via the NN method 
and the \citetalias{2017MNRAS.467.1940H} method, we construct 33 mock samples 
for each luminosity bin. For the MNN method, in order
to reduce the random noises caused by the random selection processes
involved in the method (see Section~\ref{sec:MNN} in bold letters), we
repeat the MNN method 3 times for each mock, obtaining a total of 99
samples for each luminosity bin. Our main test results for $w_p(r_p)$,
$\xi_0(s)$, $\xi_2(s)$, and $\xi_4(s)$ are presented in
Figures~\ref{fig:wp}$-$\ref{fig:xi4}
accordingly.

The detailed estimations in the figures are presented as follows. 
The correlation functions shown in the upper panels are derived as 
\begin{equation}\label{eq:X}
\overline{X} = \frac{\sum^{N_{\rm mocks}}_{i=1}X_i}{N_{\rm mocks}},
\end{equation}
where $X_i$ can be $w_p(r_p)$, $\xi_0(s)$, $\xi_2(s)$, and $\xi_4(s)$, and $i$ is the $i$th mock. 
For $X_{\rm true}$, $X_{\rm NN}$, and  $X_{\rm Hahn17}$, $N_{\rm mocks}=33$. 
For $X_{\rm MNN}$, $N_{\rm mocks}=99$. 
Error bars are the $1\sigma$ variations of $X$ computed as 
 \begin{equation}
\sigma= \sqrt{\frac{1}{N_{\rm mocks}-1} \sum^{N_{\rm mocks}}_{i=1} (X_i-\overline{X})^2}.
\end{equation}
The lower panels in Figure~\ref{fig:wp} through Figure~\ref{fig:xi4} show
the mean ratios $\overline{r_X}$ of the correlation functions.  First,
we define
\begin{equation}\label{eq:ratio}
r_{i,X}=\frac{X_i(\rm MNN/NN/Hahn)}{X_{i}({\rm true})},
\end{equation}
where $X_i$ is the same as in equation~(\ref{eq:X}). 
Then, the mean ratios are calculated as 
\begin{equation}\label{eq:mr}
\overline{r_X}=\frac{\sum^{N_{\rm mocks}}_{i=1}r_{i,X}}{N_{\rm mocks}}.
\end{equation}
The error bars in the lower panels are estimated as 
\begin{equation}\label{eq:merr}
\overline{\sigma_r}= \sqrt{\frac{1}{N_{\rm mocks}-1} \sum^{N_{\rm mocks}}_{i=1} (r_{i,X}-\overline{r_X})^2},
\end{equation}
where  $N_{\rm mocks}$ in equation~(\ref{eq:mr}) and (\ref{eq:merr})
are the same as in equation~(\ref{eq:X}).

\subsubsection{The Projected Two-point Correlation Functions} \label{sec:p2pcf}

Figure~\ref{fig:wp} shows the P2PCFs estimated with different methods for three
volume-limited mock galaxy samples. We also measure the true P2PCF 
for each sample using the true redshifts of the fiber-collided galaxies. 
The ratios between $w_p$ and $w_p(true)$ are presented
in the lower panels. To stress the essential roles of introducing the
second and the third angular neighbor galaxies in the redshift
reconstruction, we further measure the $w_p(\rm \widetilde{MNN})$ and
derive the ratio of $w_p(\rm \widetilde{MNN})$/$w_p(true)$ as shown in
the yellow curves in the lower panels. In the $\rm \widetilde{MNN}$ method, we
remove the step 3 and step 4 of the MNN method, which means only the subsample 
$\mathbf{ \Phi^{01}}$ is used in the redshift recovery process,\footnote{The $\rm \widetilde{MNN}$
  method is still different from the \citetalias{2017MNRAS.467.1940H} method.}
resulting in $29\%$ fiber-collided galaxies having
$\widetilde{d^{01}}_{\rm LOS} > 20~\mpch$.  It also implies that 
about $29\%$ of the fiber-collided galaxies are {\it randomly} distributed in the
foreground or background. Although the final distributions of
$\widetilde{d^{01}}_{\rm LOS}$ for the fiber-collided galaxies totally
trace the pair distribution $d^{01}_{\rm LOS}$ of the fiber-collision-free galaxies with known redshifts, our
key assumption of the MNN method is not adopted in the $\rm \widetilde{MNN}$ method.
As a result, it leads to obvious declinations below the
fiber-collision scale in $w_p(\rm \widetilde{MNN})$ for all three
cases as shown in the figures, and these underestimations also
extend to the intermediate scale and even the large scale, producing a
$\sim 3\%$ bias. Compared with the previous two methods and the 
$\rm \widetilde{MNN}$ method, the P2PCFs estimated from the MNN method give
the best agreement with $w_p(true)$ on all concerned scales in all
three luminosity samples. By further accounting for the pair
distributions of $d^{02}_{\rm LOS}$ and $d^{03}_{\rm LOS}$, our MNN method
successfully reduces the bias to $\sim 1\%$ on all scales.
The underestimation of $w_p$ by the $\rm \widetilde{MNN}$ method on large
scales is also corrected by the MNN method properly. However, there is a
weak luminosity dependence of the correction that can be seen in the
figures. Specifically, the MNN method tends to work better
for faint samples. For the brightest galaxy samples, the correction
on $\lesssim 1~\mpch$ scales is only slightly better than 
the $\rm \widetilde{MNN}$ method, given the larger error bars. 
We also note that there is a very small ($< 1\%$) deficiency in $w_p(\rm MNN)$ 
on large scales, indicating that these massive galaxies might have more compact
distributions around their nearest angular neighbors.
 
\subsubsection{The Redshift-space Correlation Functions} \label{sec:xi}

\begin{figure*}
\gridline{\fig{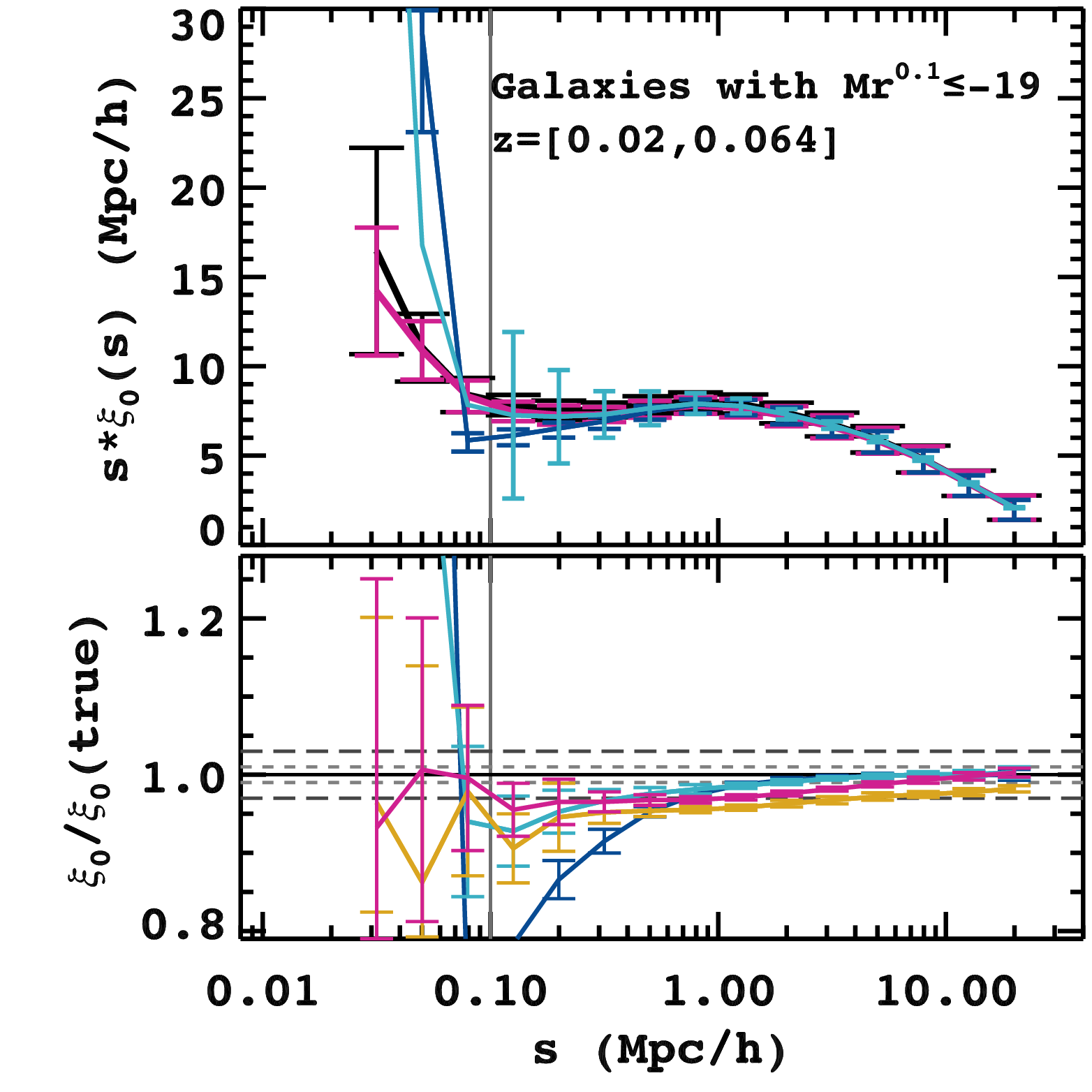}{0.33\textwidth}{(1)}
              \fig{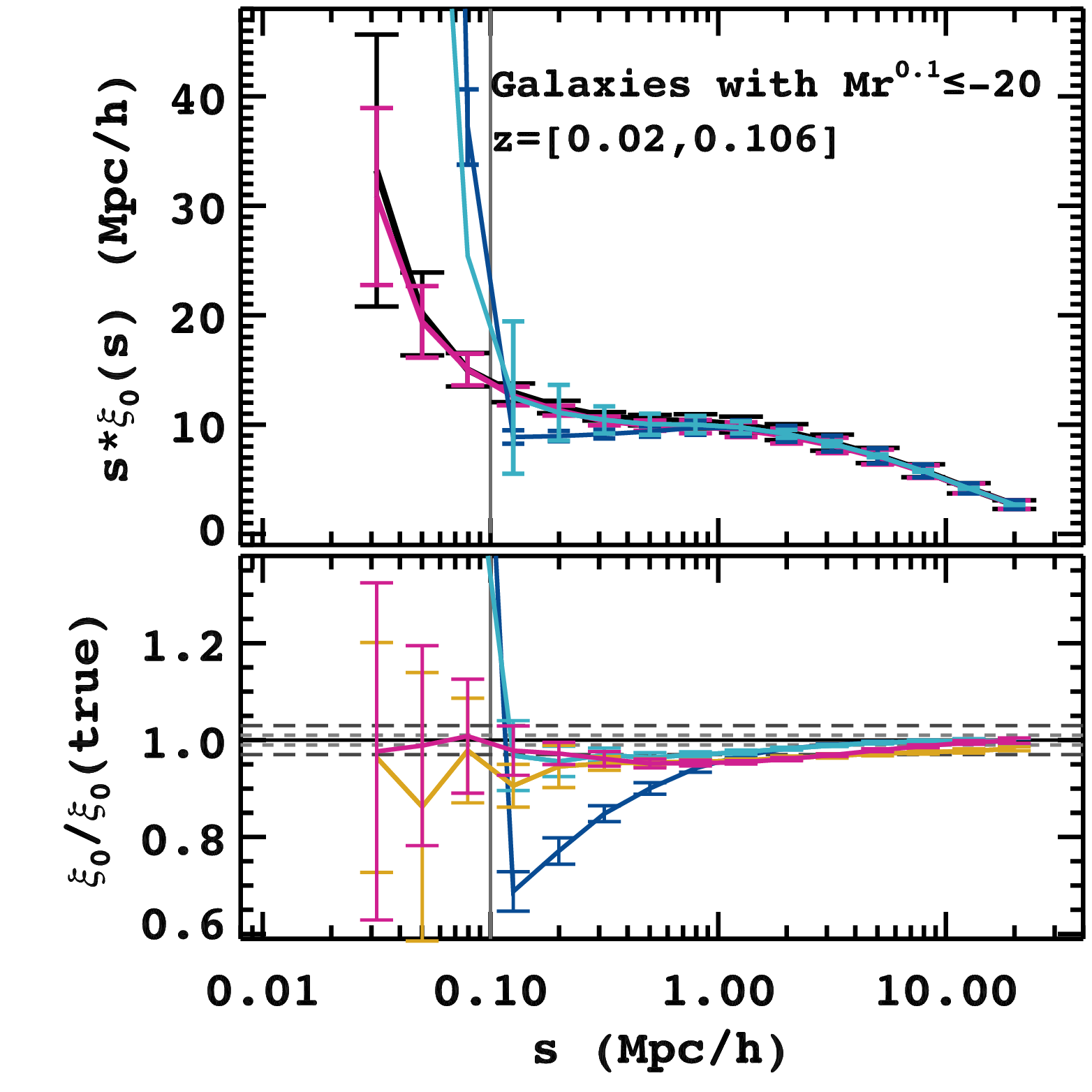}{0.33\textwidth}{(2)}
              \fig{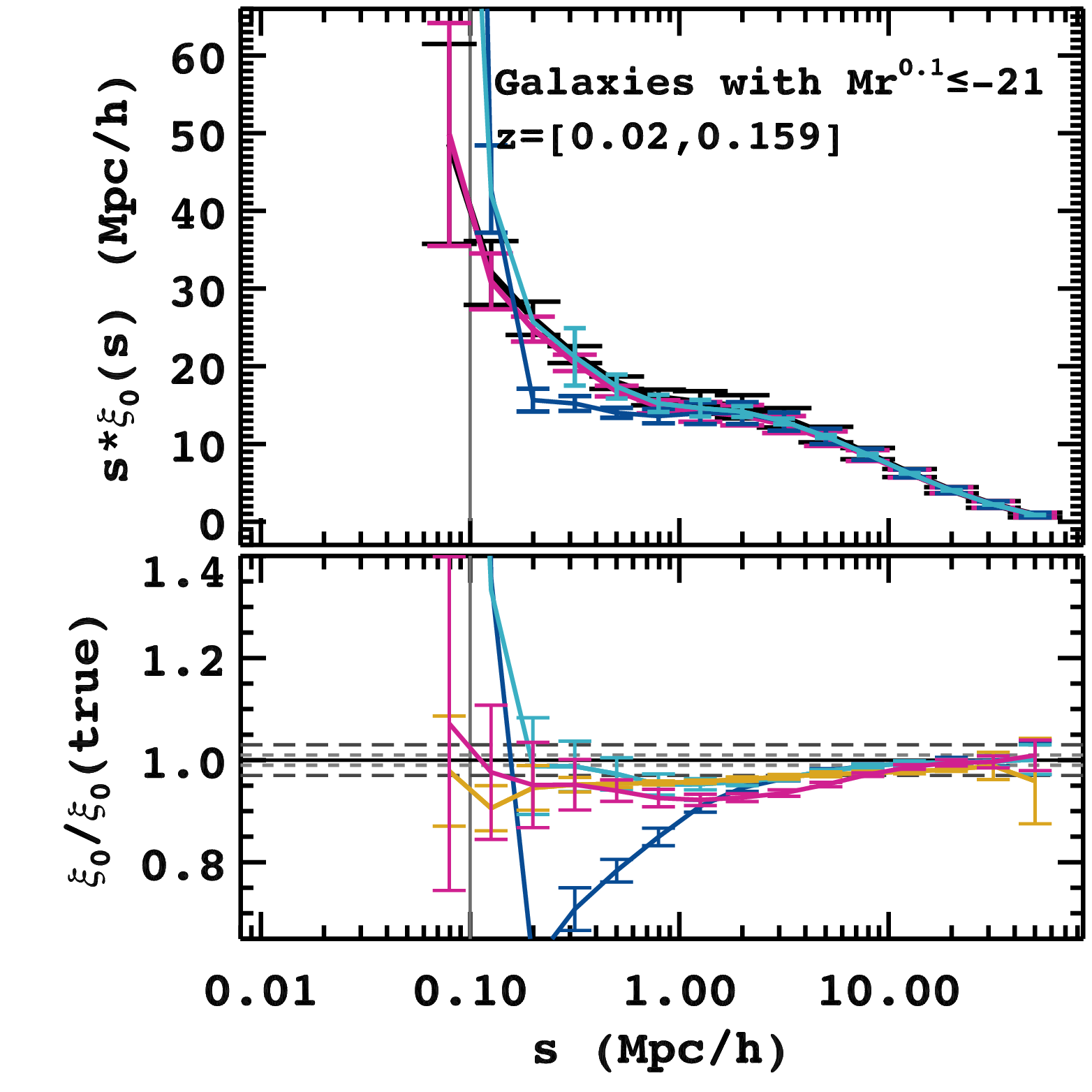}{0.33\textwidth}{(3)}
          }
          
\caption{Same as Figure~\ref{fig:wp} but for the monopole moment $\xi_0(s)$ of the redshift space correlation functions.
  \label{fig:xi}}
\end{figure*}

\begin{figure*}
\gridline{\fig{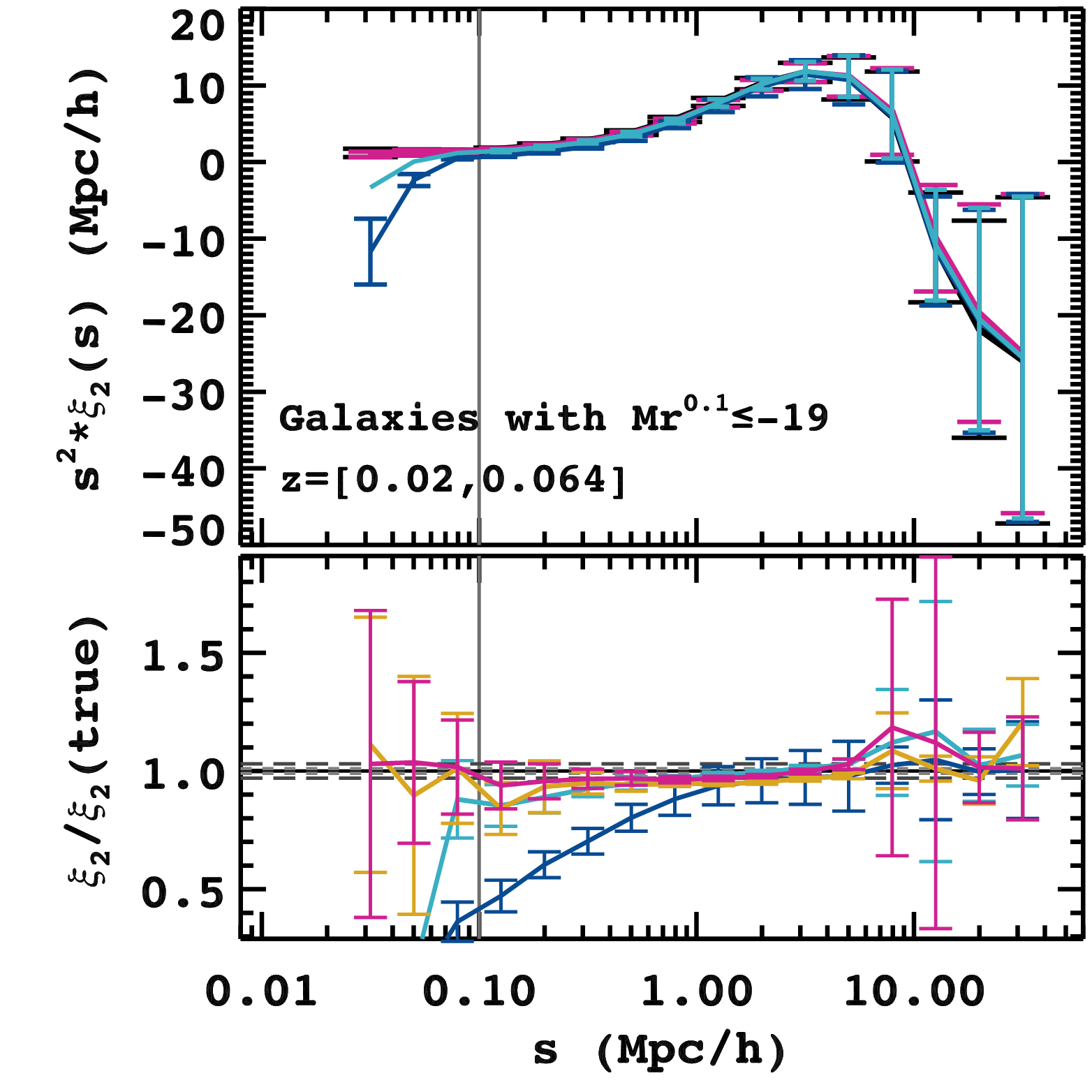}{0.33\textwidth}{(1)}
              \fig{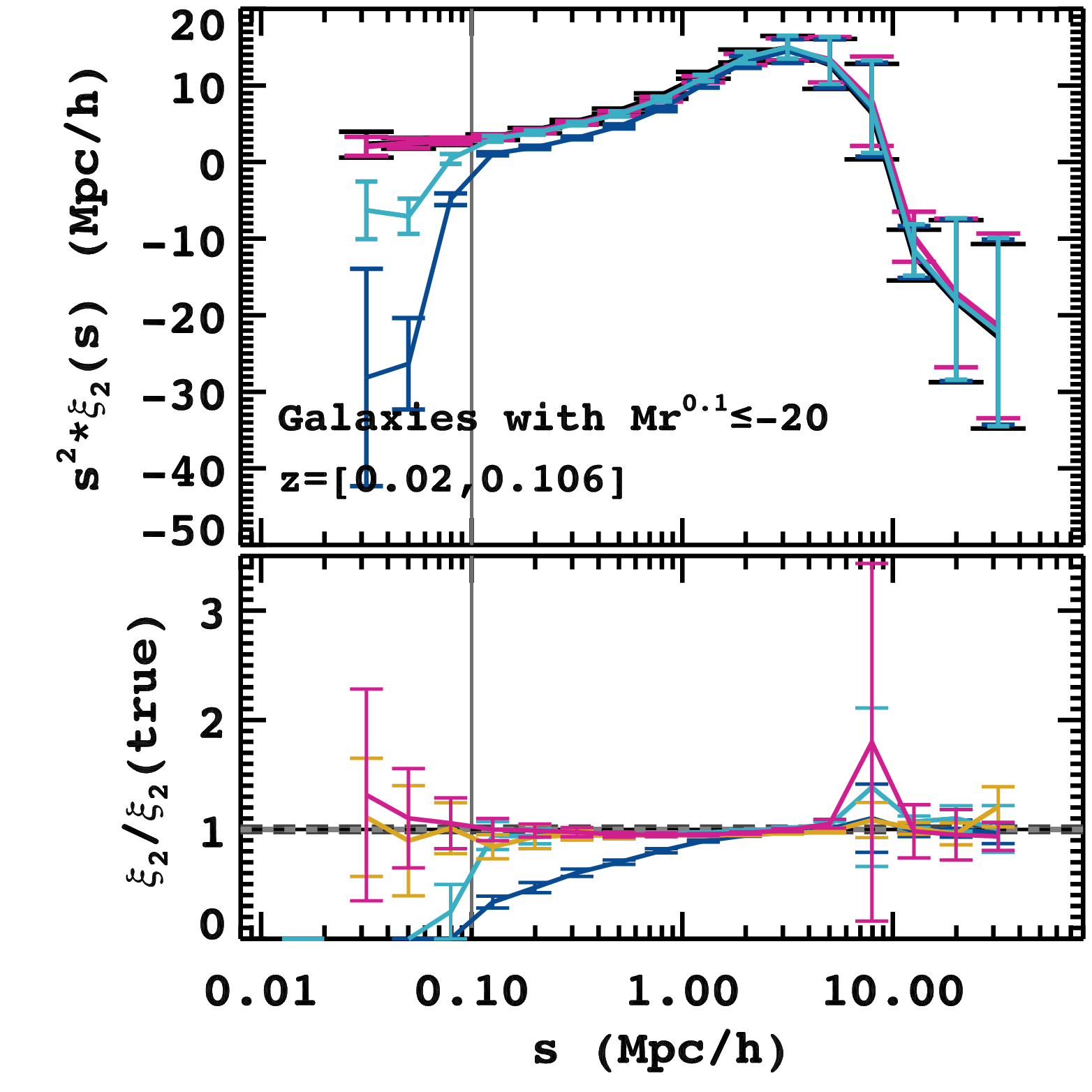}{0.33\textwidth}{(2)}
              \fig{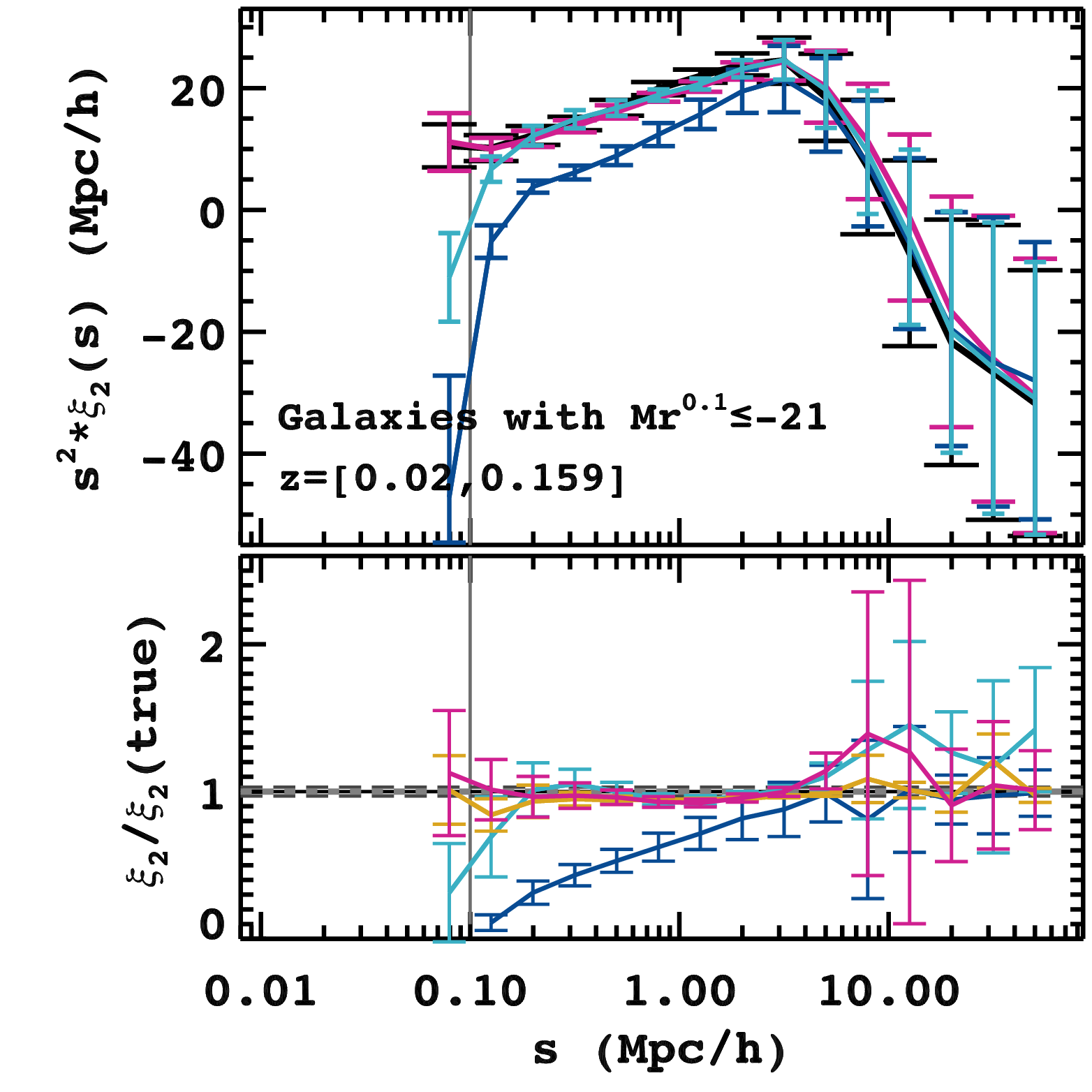}{0.33\textwidth}{(3)}
          }
\caption{Same as Figure~\ref{fig:wp} but for the quadrupole moment $\xi_2(s)$ of the redshift space correlation functions.
              \label{fig:xi2}}
\end{figure*}

\begin{figure*}
\gridline{\fig{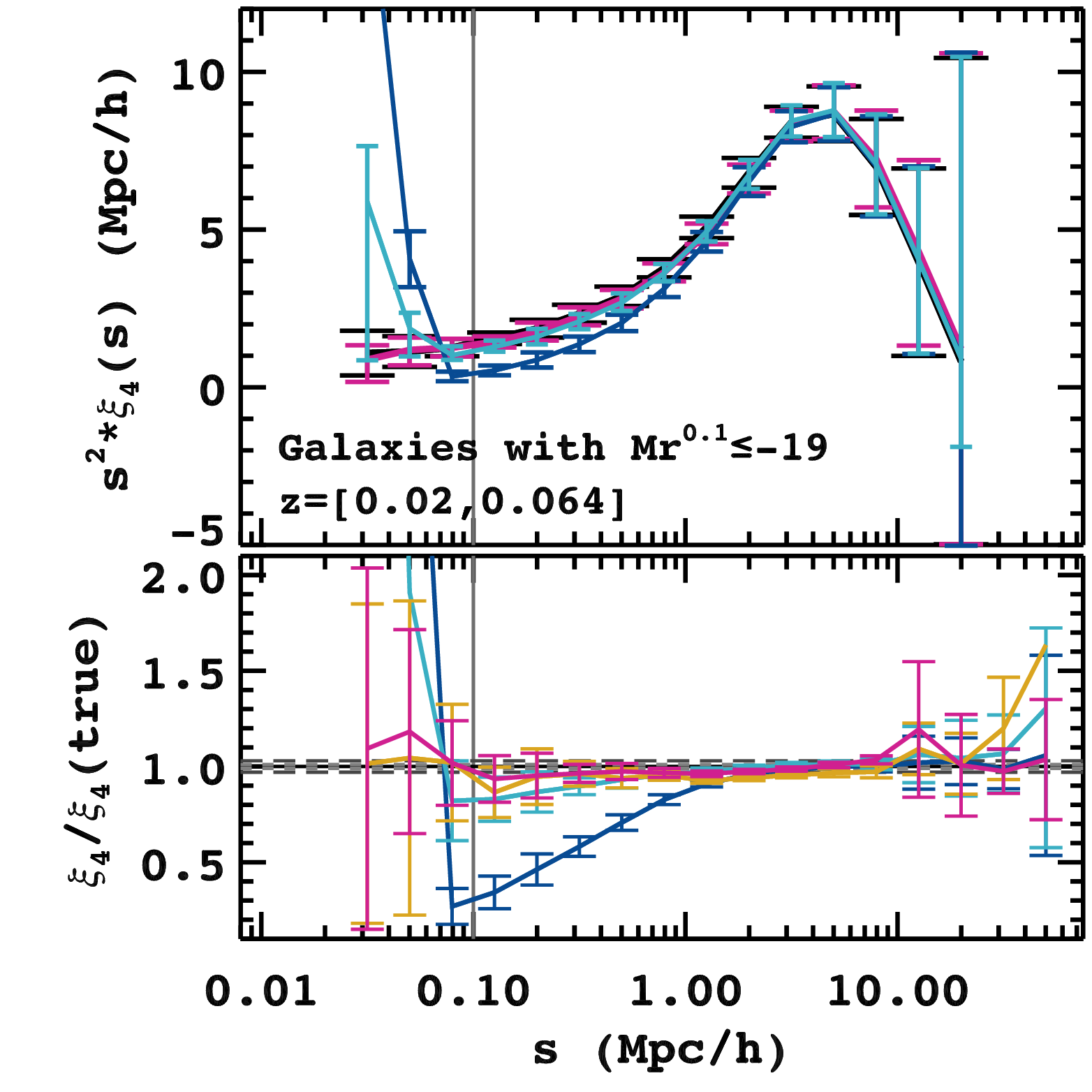}{0.33\textwidth}{(1)}
              \fig{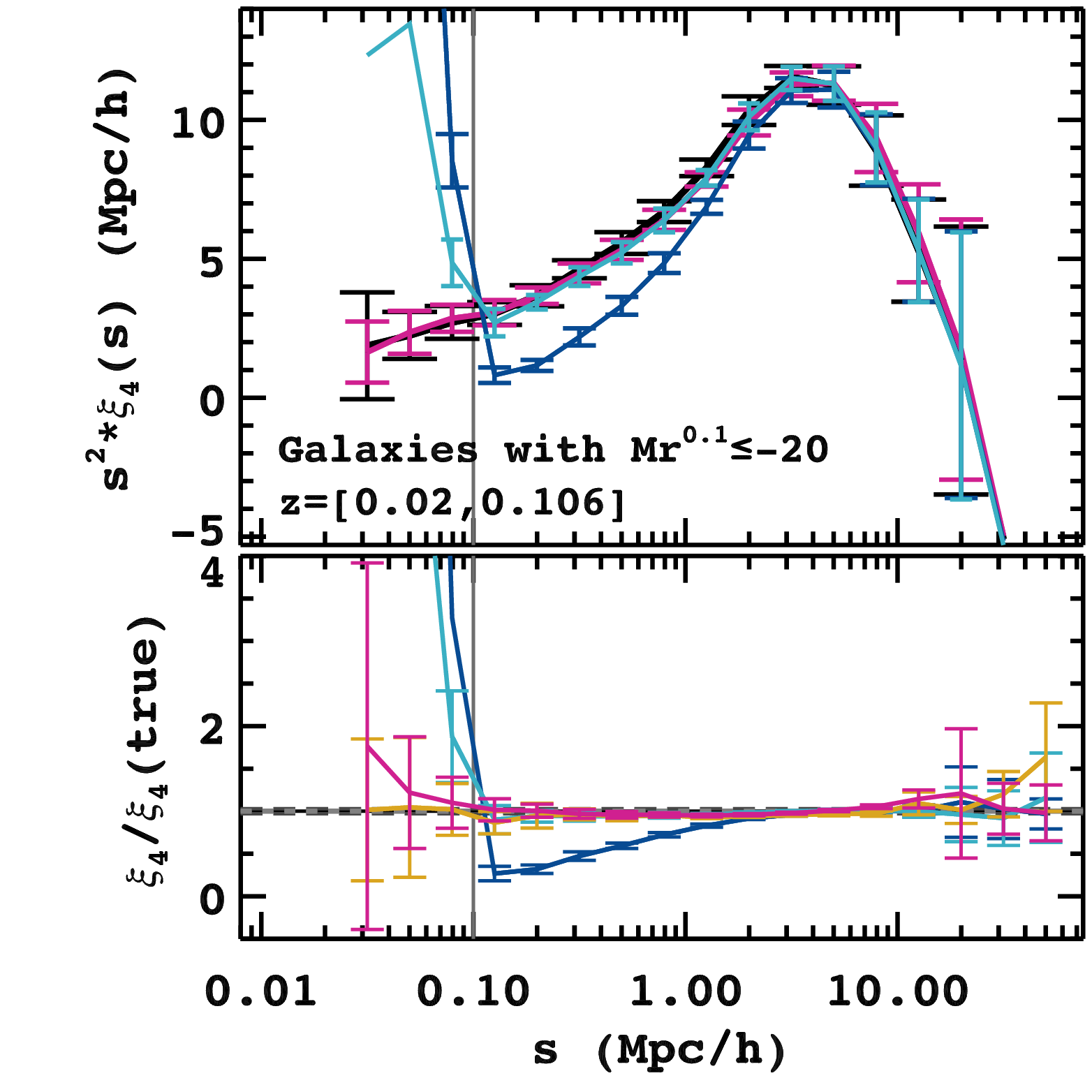}{0.33\textwidth}{(2)}
              \fig{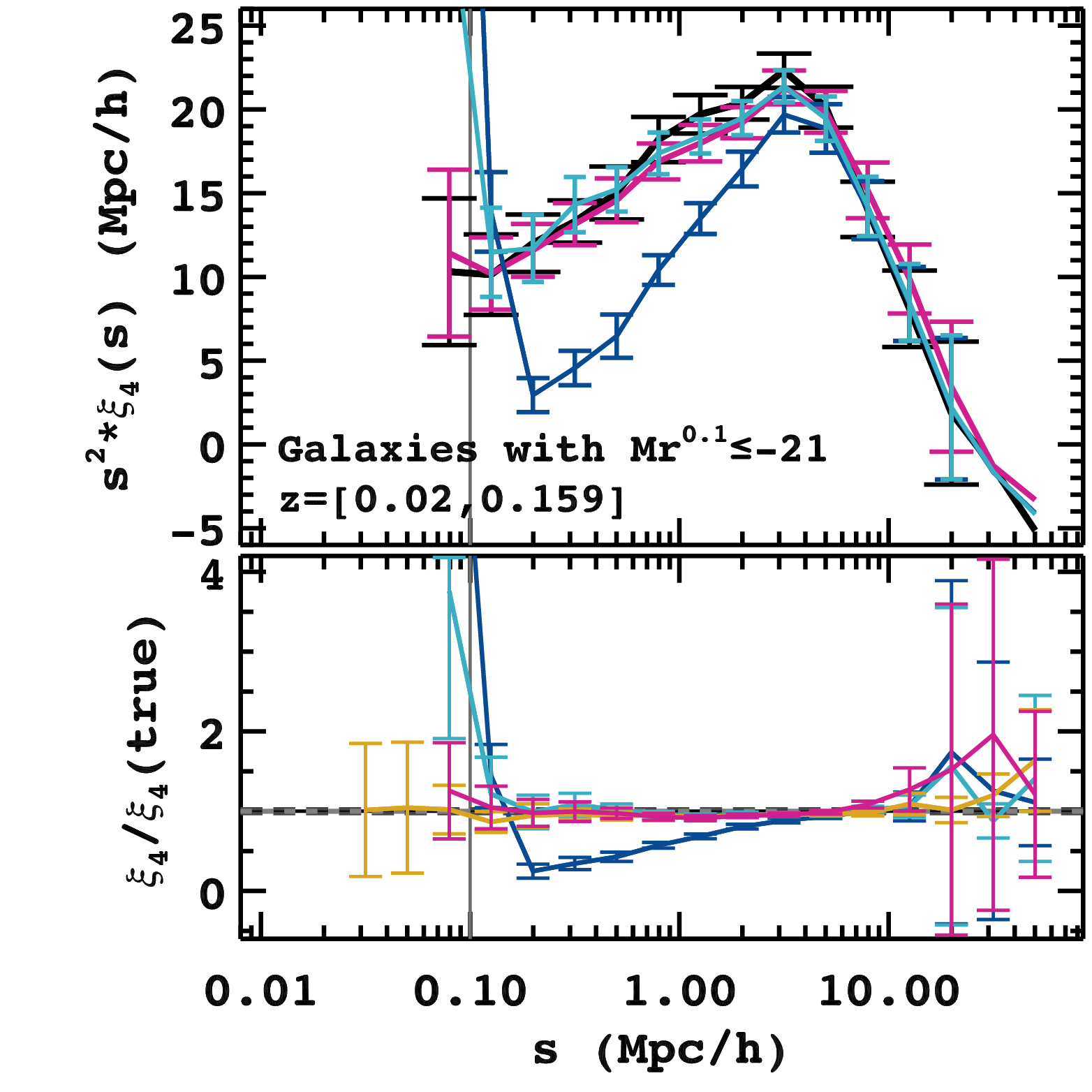}{0.33\textwidth}{(3)}
          }
\caption{Same as Figure~\ref{fig:wp} but for the hexadecapole moment $\xi_4(s)$ of the redshift space correlation functions.
              \label{fig:xi4}}
\end{figure*}

The non-zero multipole moments of the redshift-space correlation
functions $\xi(s,\mu)$ are presented in Figure~\ref{fig:xi} for
$\xi_0(s)$, Figure~\ref{fig:xi2} for $\xi_2(s)$, and
Figure~\ref{fig:xi4} for $\xi_4(s)$. The symbols and colors are the
same as those in Figure~\ref{fig:wp} for the P2PCFs. Below the fiber-collision
scale, the MNN method apparently provides a better correction compared
with other methods. The $\xi_0(s)$s estimated from the NN method and 
the \citetalias{2017MNRAS.467.1940H} method exhibit significant deviations from the
true values, while that from the MNN method agrees with the true $\xi_0(s)$ 
within the 1$\sigma$ error. On scales between 0.1 and 1.0 $\mpch$, 
we see significant deviations from the true $\xi_0(s)$ in the NN
method, while the deviations in the MNN and the \citetalias{2017MNRAS.467.1940H}
methods are quite small. Although the MNN method gives a better recovery 
of the true $\xi_0(s)$, we note that there is an underestimation of 
$3\% \sim 10\%$ beyond the fiber-collision scale and extending to $\sim$ 2.0$\mpch$. 
The underestimations can also be seen in $\xi_2(s)$, $\xi_4(s)$, but is absent in $w_p$. 
A simple explanation is that these underestimations 
are mainly caused by the linear coherent motion, 
or the peculiar velocities of galaxies. 
The influence of peculiar velocities on galaxy clustering has been well 
averaged out through the line-of-sight integration for the projected two-point correlation functions. 
In the case of the redshift space correlation functions, however, the galaxy 
peculiar velocities play a crucial role on the small scales, 
especially on the one-halo scale. Therefore, in all of the fiber-collision 
correction methods, the effects of the peculiar velocities should be further corrected in the ideal case.

Figure~\ref{fig:xi2} and Figure~\ref{fig:xi4} show the
redshift-space correlation functions $\xi_2(s)$ and $\xi_4(s)$.  The
overall performances of the different fiber-collision
correction methods for these measurements are quite similar to those for $\xi_0(s)$, 
except that the reduction seen on intermediate scales in $\xi_0(s)$ is not prominent
at all in the recovery of $\xi_2(s)$ and $\xi_4(s)$.

\section{Discussions}\label{sec:diss}
\subsection{Limitations of the MNN Method}\label{sec:limit}

Although the MNN method produces the best results 
around the fiber-collision scale, its application strongly relies on the 
availability of the collision-free galaxies with measured redshifts within the fiber-collision scale. 
These collision-free galaxies are normally only available for surveys with overlapping tiling regions. 
Moreover, the existence of biases indicates that there is still room to further improve the MNN 
method.

The key assumption of the MNN method is that
every fiber-collided galaxy is assumed to be in association with its
three nearest angular neighbors, i.e., they reside in the same
large-scale environment. This assumption guarantees that not only
the distribution of $\widetilde{d^{01}}_{\rm LOS}$ follows the distribution of 
$d^{01}_{\rm LOS}$ derived from galaxies with measured redshifts, 
but $\widetilde{d^{02}}_{\rm LOS}$ and
$\widetilde{d^{03}}_{\rm LOS}$ also approximately trace the distribution of
$d^{02}_{\rm LOS}$ and $d^{03}_{\rm LOS}$. This assumption has a solid strong 
statistical basis as shown in Table~\ref{tab:dstatistics} and Figure~\ref{fig:hist}, 
which are derived from both the observational data and the corresponding mocks. 
The basis also tightly depends on a precondition, 
that there must be galaxies with measured redshifts below the fiber-collision scale. 
Actually, it is not too difficult to achieve this precondition. For a single-pass survey, 
there are often some overlapping tiling regions that enable some galaxy pairs within 
the fiber-collision scale to be both observed, while multi-pass surveys can provide 
plenty of these kinds of galaxies. With more galaxies observed spectroscopically below 
the fiber-collision scale, the pair distributions are closer to the real distributions, and the clustering 
measured by applying the MNN method becomes more reliable and robust. 
This argument works for any methods that are trying to correct 
the missing redshifts in two-point statistics.

The results of the MNN method exhibit a slight luminosity dependence, 
particularly the faint samples show a better correction below the fiber-collision scale 
than the luminous samples. This trend could be explained by the assumption of 
the MNN method as mentioned before. Fiber collisions mainly happen in galaxy-dense 
regions such as galaxy clusters, where there tend to be more brighter galaxies 
than seen in less dense regions \citep{1980ApJ...236..351D,1983ApJ...267..465D,2018arXiv180403097W}. 
These luminous galaxies also have a stronger clustering strength
than faint galaxies \citep{2005ApJ...630....1Z,2011ApJ...736...59Z}.
In future works, we plan to further consider this dependence on the galaxy luminosity 
to avoid the limitations arising from the indiscriminate treatments to the different luminous populations.

\subsection{Differences from Previous Methods}
For the NN method, the assumption that each collided galaxy is associated with 
its nearest neighbor is too strong to recover the true distributions of the fiber-collided populations. 
This leads to an overclustering bump below the fiber-collision scale, as shown in
Figure \ref{fig:wp}. We argue that the satisfactory agreement on
intermediate scales of $w_p$(NN) is actually a pair compensation effect
that only works for $w_p$. When it comes to the multipole moments of the correlation functions,
the drawbacks are apparent: extremely high biases are shown 
on small scales and even extending to the intermediate scales, as shown 
in Figure \ref{fig:xi} through \ref{fig:xi4}. These extreme behaviors
are directly caused by insufficient modeling of the intrinsic separations of galaxy pairs 
\citep{1972MNRAS.156P...1J, 1987Natur.327..210P,
  1988ASPC....4..257H, 1992ApJ...385L...5H}.  More specifically, 
panel (1) in Figure \ref{fig:hist} is modeled by a $\delta$ function in the 
case of the NN method. In reality, the line-of-sight separations
of the nearest pairs display a Gaussian-like distribution with a higher
and broadening wing, as shown in Figure \ref{fig:hist}. A coherence 
length around 20 $\mpch$ can also be clearly seen \citepalias{2017MNRAS.467.1940H}.  
This coherence length is understandable if we recall that a 
typical galaxy coherence length of 50 $\mpch$ is measured in the local
peculiar velocity field \citep{1987Natur.327..210P}. Since the galaxies that are most likely affected 
by the fiber-collision effect usually reside in dense regions of the universe, 
where the impact of the peculiar velocities are relatively stronger than those in less 
dense regions, their coherence length is thus smaller.

To alleviate the problem caused by the intrinsic velocity dispersion of galaxy pairs, 
\citetalias{2017MNRAS.467.1940H} conservatively added a Gaussian scatter to 
the pair displacements for about 70\% of pairs within $3\sigma_{\rm LOS}$. 
The remaining 30\% of pairs are kept the same as in the NN method.  As expected, the biases 
of the power spectrum are improved significantly compared with the results of the NN 
method (see their Figure 3 and Figure 6), despite the correction still being severely limited, 
especially for the quadrupole power spectrum. In our test of their method, the correlation function 
appears to have been corrected to a better degree. This might have beeb caused by many subtle differences 
between our implementation and their original ones. First of all, we apply 
the \citetalias{2017MNRAS.467.1940H} method in real space, while their implementation 
is in Fourier space. There are also a few differences in the galaxy samples used by our work and theirs. 
For example, we use the SDSS DR7 \full sample, which covers a smaller volume compared with 
their BOSS DR12 CMASS sample. The fiber-collision scales are also different for the two SDSS samples. 
The simulation we adopt to build the mocks has a higher resolution than theirs. 
So it is not too surprising that the \citetalias{2017MNRAS.467.1940H} method 
works well in our tests, which give a slightly better correction results than the NN method.  
Lastly, we are also aware that one method may give different correction results for 
the correlation functions in real space and for the power spectrum in Fourier space. 
However, further exploring the different performances of a method in different spaces is beyond the scope of this paper.

In the MNN method, first, we adopt a fixed correlation length of 20 $\mpch$, 
rather than the $3\sigma_{\rm LOS}$ scatter applied by \citetalias{2017MNRAS.467.1940H}. 
Second, two more angular neighbors are taken into account mainly 
based on the statistical fraction distributions of galaxy pairs from the fiber-collision-free 
galaxies. Third, in step 2, instead of randomly selecting a galaxy from $N$ galaxies 
in the same $\Delta\theta$ bin of $\mathbf{ \Phi^{01} }$, we introduce a parameter $N_{\rm near}$ and we only select 
a galaxy from $N_{\rm near}$ neighbors of the galaxy ``$\widetilde{1}$.'' The function of 
$N_{\rm near}$ is to set a range of the possible comoving distances or redshifts. 
We have tested that a blind and random selection of the $N$ galaxies in $\Delta\theta$ bin 
can result in an underestimation of clustering 
below the fiber-collision scale. However, the small-scale clustering is not very sensitive 
to the choice of $N_{\rm near}$ as shown in Figure~\ref{fig:cf2params}.
As expected, after these improvements the MNN method effectively lowers the
overclustering bump of the two-point correlation functions below the fiber-collision scale. 
It also presents superior measurements on intermediate scales compared with other methods. 
For $w_p$, the biases are reduced to $1\%$, and the small
deviations are well under the $1\sigma$ measurement errors.  For $\xi_0$, although the underestimation 
is as large as $\sim 10\%$ as shown in the brightest samples, this underestimation can be partly ascribed 
to the unmodeled coherent motions of the collided galaxies, a problem that we do not try to resolve in this work. 

\subsection{Choices of Model Parameters}\label{sec:testptrs}
\begin{figure*}
\gridline{\fig{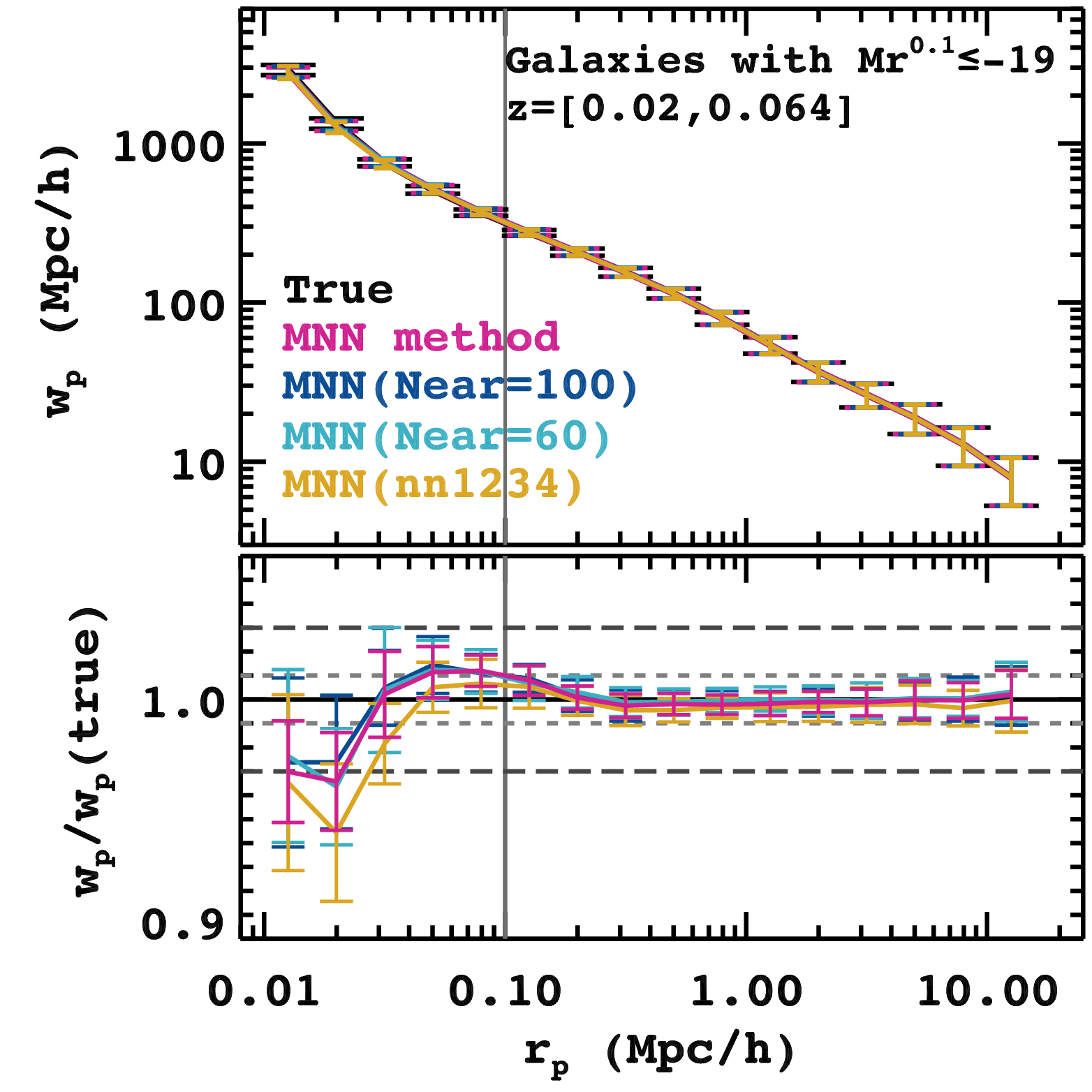}{0.4\textwidth}{(1)}
              \fig{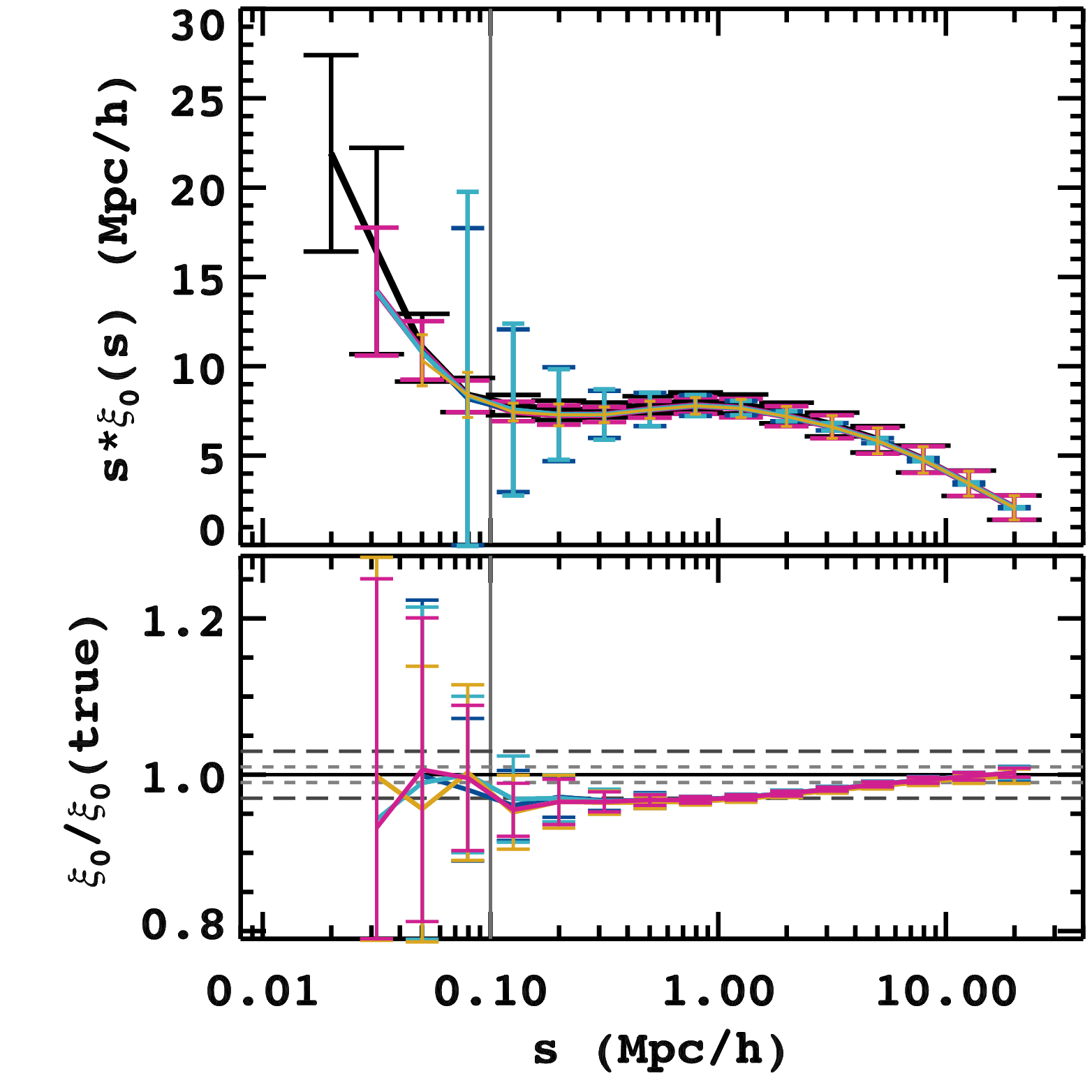}{0.4\textwidth}{(2)}
          }
\gridline{\fig{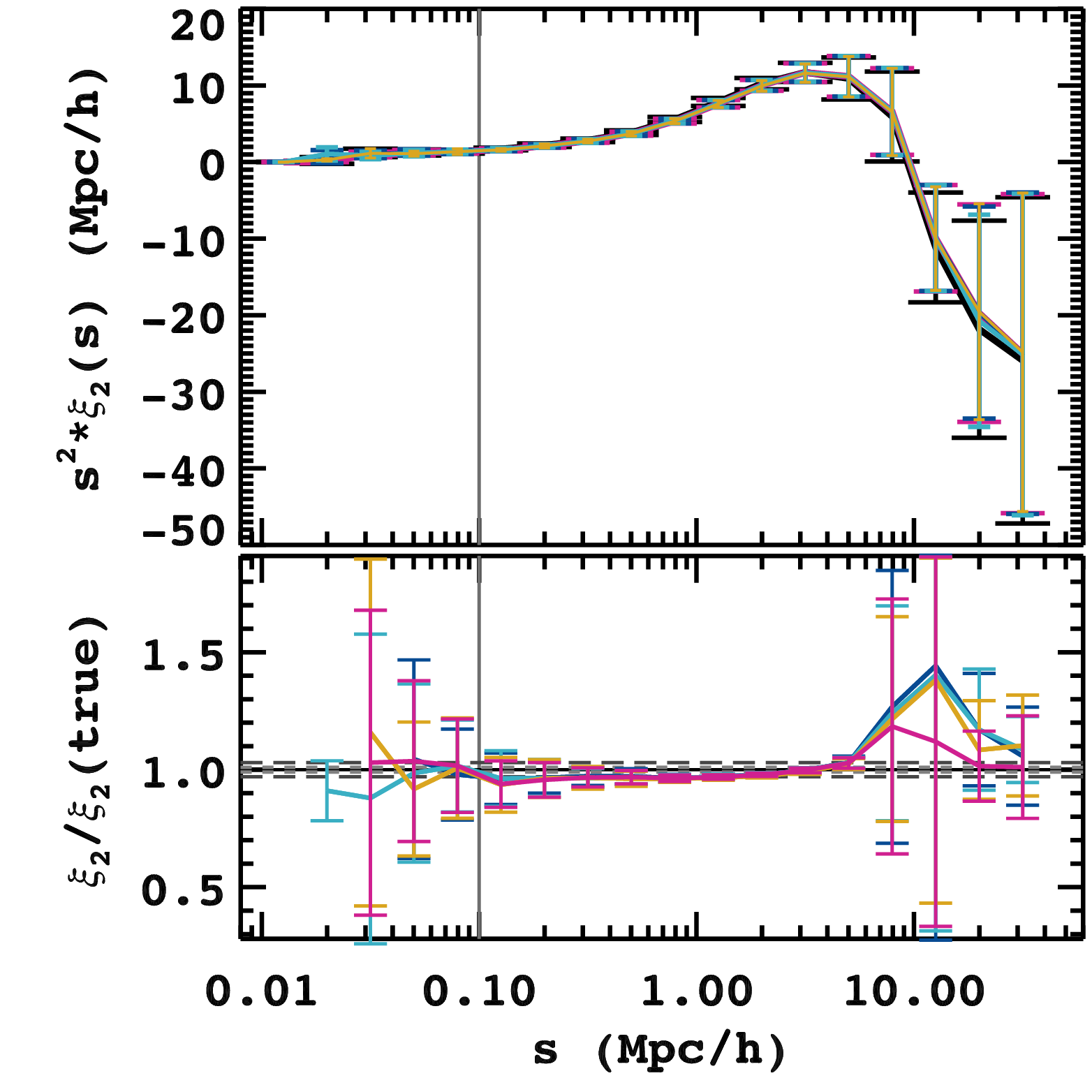}{0.4\textwidth}{(3)}
              \fig{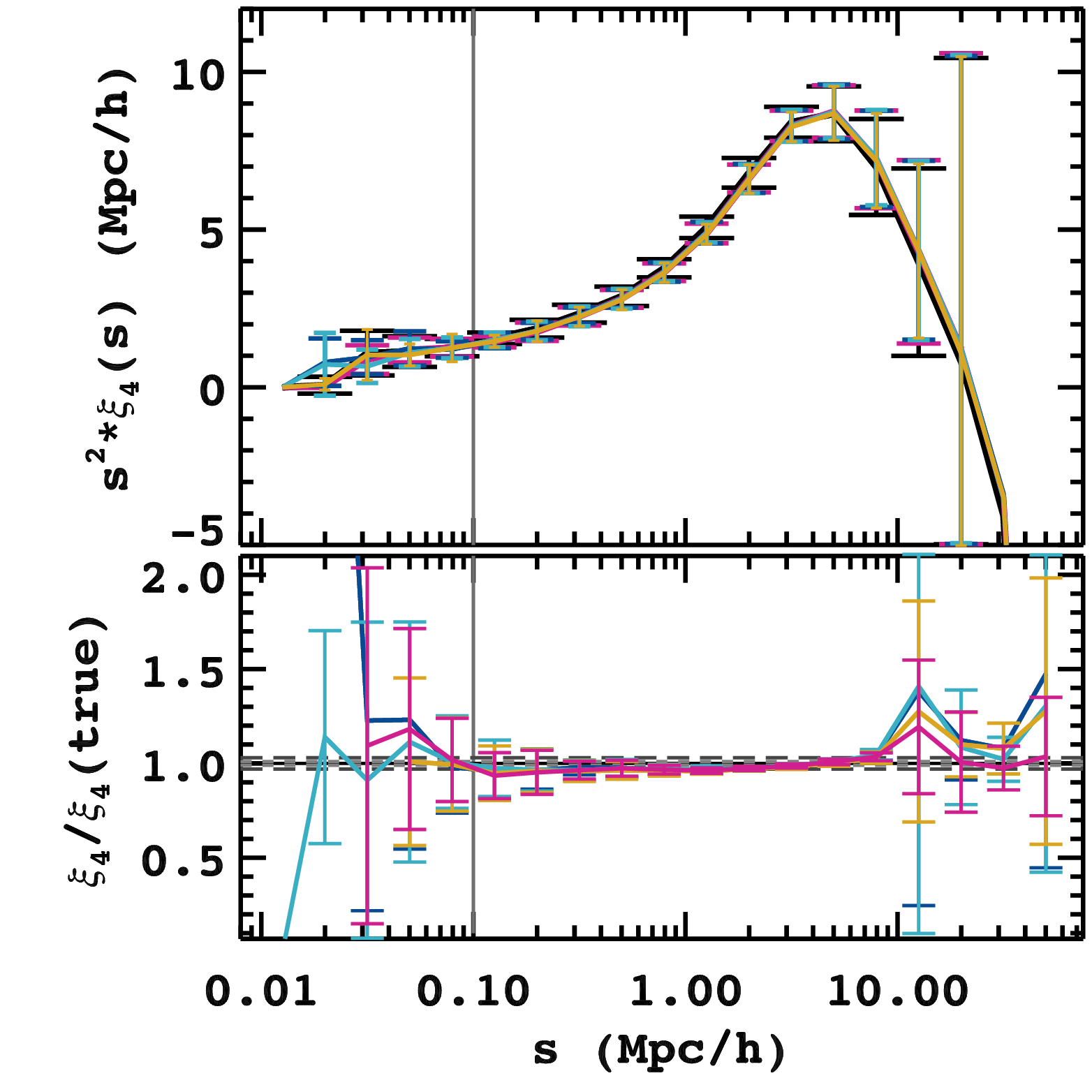}{0.4\textwidth}{(4)}
          }

\caption{Testing the performance of the MNN method with three different choices of parameters: 
(1) we include four neighbors (yellow) to perform the MNN, instead of three neighbors (magenta); 
(2) we set $N_{\rm near}=100$ (dark blue) and $N_{\rm near}=60$ (light blue) instead of the first choice of $N_{\rm near}=30$.
Panels (1), (2), (3), and (4) show comparisons of $w_p$, $\xi_0$, $\xi_2$, and $\xi_4$, respectively. 
We can see that neither the use of four neighbors 
nor the choices of $N_{\rm near}$ are very sensitive to $w_p$, $\xi_0$, $\xi_2$, and $\xi_4$. 
The gray vertical line and the horizontal dotted and dashed lines are the same as in Figure~\ref{fig:wp}.
              \label{fig:cf2params}}
\end{figure*}

To see the performance of the MNN method 
if different neighbors and different choices of $N_{\rm near}$ are used, 
we perform further tests, as shown in Figure~\ref{fig:cf2params}. We try three different cases: 
including the fourth neighbor (yellow curves in Figure~\ref{fig:cf2params}), 
setting $N_{\rm near}=60$ (light blue curves in Figure~\ref{fig:cf2params}), and 
setting $N_{\rm near}=100$ (dark blue curves in Figure~\ref{fig:cf2params}). 
We see that only a small bias of $w_p$ is aroused by using four neighbors below 
the fiber-collision scale, which is still under the measurement errors as shown. 
Actually, we have also tried the case of only including two neighbors. This leads to 
a 3\% overclustering bump of $w_p$ below the fiber-collision scale, 
similar to the result from the NN method. Therefore, as supported by 
the pair fraction statistics in Table~\ref{tab:dstatistics}, the choice of three 
neighbors gives the best estimation on small scales. On the other hand, 
the clustering is also not sensitive to the choice of $N_{\rm near}$ at all. 
Even for the multipole moments of the correlation functions, the biases arising 
from the use of different $N_{\rm near}$ in the MNN method are very small.

\section{Summary} \label{sec:summary}

In this paper, we have developed a new method to correct for the fiber-collision effect, 
which is a common problem in modern spectroscopic galaxy surveys. 
We mainly focus on correcting galaxy clustering 
below the fiber-collision scale and the intermediate scale $\lesssim 10\mpch$. 
The MNN method is basically built upon the previously 
proposed NN method and the \citetalias{2017MNRAS.467.1940H} method. 
The key assumption of this method is that the
fiber-collided galaxy is in association with its three nearest angular
neighbors. By statistically investigating the line-of-sight comoving
separations of the neighboring galaxy pairs with resolved
spectroscopic redshifts, we find the association length is $\sim 20\mpch$. 
To test the method, we use a high resolution $N$-body simulation to construct 
33 mock galaxy catalogs mimicking the observational selection of the SDSS DR7 \full\ sample 
from NYU-VAGC. Our main tests of the MNN method are 
performed with these 33 mocks.  
By comparing the projected two-point correlation functions and the multipole
moments of the correlation functions in real space for three different
volume-limited luminosity threshold samples, we demonstrate that the MNN
method can reduce the bias to $1\%$ for $w_p$, which is a significant 
improvement compared with other methods.

The advantages of the MNN method are as follows. 
First, this method is built upon the intrinsic distribution of galaxy pairs. 
This distribution is recovered from galaxy pairs within the fiber-collision scale 
that still have measured redshifts, thanks to overlapping tiling regions in most spectroscopic surveys.
Second, a better estimation of the galaxy two-point statistics
can be attained below the fiber-collision scale and on the intermediate scale compared 
with other methods. Third, because the MNN method can assign new redshifts to fiber-collided galaxies 
and have a good recovery of the true redshift distribution, in principle it can also measure 
the power spectrum accurately. We will test the performance of our method in Fourier space in our future work. 
We have also summarized the limitations of the MNN method in Section \ref{sec:limit}, 
where we believe the coherent motion of galaxy pairs is quite important in the recovery of 
the redshift-space correlation functions. We are working to improve the MNN method, 
and will test the improved method with newer data in future works.

\acknowledgments{ACKNOWLEDGMENTS}\\

We are grateful to the anonymous referee for insightful suggestions
that significantly improved this paper.
L.Y. thanks Hong Guo and Zhigang Li  and Feng Shi for their useful
conversations on mock construction and galaxy clustering
measurements. The work is supported by the 973 Program
(Nos. 2015CB857002, 2015CB857003) and NSFC (11320101002, 11533006, \&
11621303). J.H. is supported by JSPS Grant-in-Aid for Scientific
Research JP17K14271. Kavli IPMU is supported by World Premier International Research Center Initiative (WPI), MEXT, Japan. 

\vspace{5mm}

\bibliography{mnn}

\end{document}